\documentclass[11pt]{article}
\usepackage{jheppub}
\usepackage{graphicx}
\usepackage{color}
\usepackage[dvipsnames]{xcolor}

\newcommand{\gev}{{\rm GeV}}
\newcommand{\tev}{{\rm TeV}}

\title{Forward production of prompt neutrinos from charm  in the atmosphere and
at high energy colliders}

\author[a]{Weidong Bai,}
\author[b]{Milind Diwan,}
\author[c]{Maria Vittoria Garzelli,}
\author[d]{Yu Seon Jeong,}
\author[b,e,1]{\\Karan Kumar,
\note{now at: Department of Physics, Columbia University, 116th and Broadway, New York, NY, USA, 10027}}
\author[f]{Mary Hall Reno}

\affiliation[a]{School of Physics, Sun Yat-sen University, 
Guangzhou, Guangdong 510275, P. R. China}
\affiliation[b]{Brookhaven National Laboratory, Upton, New York, USA}
\affiliation[c]{II Institut f\"ur Theoretische Physik, Universit\"at Hamburg\\
Luruper Chaussee 149, D-22761, Hamburg, Germany}
\affiliation[d]{High Energy Physics Center, Chung-Ang University,
Dongjak-gu, Seoul 06974, Republic of Korea}
\affiliation[e]{Department of Physics and Astronomy, Stony Brook University, Stony Brook, NY 11794, USA}
\affiliation[f]{Department of Physics and Astronomy, University of Iowa, Iowa City, IA 52242, USA}

\emailAdd{baiwd3@mail.sysu.edu.cn}
\emailAdd{diwan@bnl.gov}
\emailAdd{maria.vittoria.garzelli@desy.de}
\emailAdd{yusjeong@cau.ac.kr}
\emailAdd{fk237@cornell.edu}
\emailAdd{mary-hall-reno@uiowa.edu}

\abstract{
The high-energy atmospheric neutrino flux is dominated by neutrinos from the decays of charmed hadrons produced in the forward direction by cosmic ray interactions with air nuclei. We evaluate the charm contributions to the prompt atmospheric neutrino flux as a function of the center-of-mass energy $\sqrt{s}$ of the hadronic collision and of the center-of-mass rapidity $y$ of the produced charm hadron. 
Uncertainties associated with parton distribution functions are also evaluated as a function of $y$.
We find that the $y$ coverage of LHCb for forward heavy-flavour production, 
complemented by the angular coverage of present and future forward neutrino experiments at the LHC, bracket the most interesting $y$ regions for the prompt atmospheric neutrino flux. 
At $\sqrt{s}=14$ TeV foreseen for the HL-LHC phase,  nucleon collisions in air contribute to the prompt neutrino flux  prominently below $E_\nu\sim 10^7$~GeV. Measurements of forward charm and/or forward neutrinos produced in hadron collisions up to $\sqrt{s}=100$ TeV, which might become possible at the FCC, are relevant for the prompt atmospheric neutrino flux up to $E_\nu=10^8$ GeV and beyond.
}

\begin{document}

\hfill {\tt BNL-224879-2023-JAAM}

\maketitle

\section{Introduction}

Through the detection of neutrinos produced in the Sun, in the atmosphere, in reactors and in laboratory accelerator beams, information concerning neutrino properties has been inferred with increasing precision over the years \cite{Giunti:2022aea,Super-Kamiokande:2001bfk,SNO:2002tuh}. 
Ongoing and future experiments that use 
these neutrinos will refine some of the existing measurements, provide new ones, expand our knowledge of the Standard Model (SM) of interactions and probe parameter spaces of new physics beyond the Standard Model (BSM) \cite{Huber:2022lpm}. 
Data on neutrino interactions are useful to constrain the partonic structure of nucleons and nuclei.
Direct neutrino cross-section measurements 
using laboratory neutrino beams 
have been made for neutrino energies up to $E_\nu=360$ GeV~\cite{Workman:2022ynf}. %Neutrino cross-section measurements extracted from 
Energy and angular distributions of neutrino events in the IceCube detector have allowed to extend neutrino cross-section measurements to $E_\nu\sim 1$~PeV, however, with large uncertainties \cite{IceCube:2017roe,IceCube:2020rnc,Bustamante:2017xuy}, and projections to higher energies have been made \cite{Valera:2022ylt,Esteban:2022uuw}.
On the other hand, programs for studying long-baseline neutrino oscillation physics 
with laboratory neutrino beams 
focus on neutrino beam average energies up to 17~GeV \cite{Workman:2022ynf}. Oscillation physics has been part of the research program of neutrino telescopes \cite{IceCube:2014flw,KM3Net:2016zxf} and other underground detectors~\cite{Super-Kamiokande:2001bfk,SNO:2002tuh} continuing into the future \cite{Hyper-Kamiokande:2018ofw,DUNE:2020lwj,DUNE:2021tad,JUNO:2022mxj}.

The highest energy neutrinos in the laboratory are produced at hadron colliders. As already discussed in the 1980's and 1990's,  there is a large flux of  neutrinos that goes in the very forward direction \cite{DeRujula:1984pg,Winter:1990ry,DeRujula:1992sn,Vannucci:1993ud}. The highest energy electron and tau (anti)neutrinos
in the forward region come from the production and decays of heavy-flavour hadrons, especially charm hadrons \cite{Bai:2020ukz, Bai:2021ira,Bai:2022jcs,Kling:2021gos}.
Over the past few years, two experiments at the Large Hadron Collider (LHC), FASER$\nu$ \cite{FASER:2019dxq,FASER:2022hcn} and SND@LHC \cite{SHiP:2020sos,SNDLHC:2022ihg}, 
have been prepared to detect such neutrinos with energies up to a few TeV and investigate  related physics.
These experiments are installed in existing service tunnels and have started collecting data during Run 3 at the LHC. First observations of collider neutrinos by these collaborations have been reported \cite{FASER:2023zcr,SNDLHC:2023pun}. Already during the LHC Run 2, a prototype detector for FASER$\nu$ recorded the first few candidate neutrino events %during the LHC Run 2 
\cite{FASER:2021mtu}.
A next-generation of forward neutrino experiments for the High-Luminosity runs of the LHC (HL-LHC) has been recently proposed. Most of them could be located into a purpose-built Forward Physics Facility (FPF) \cite{Anchordoqui:2021ghd,Feng:2022inv}, 
but further solutions, involving other LHC areas and neutrino pseudo-rapidities, $\eta_\nu=-\ln\tan(\theta/2)$ where $\theta$ is the angle relative to the beam direction, are also under investigation.
Neutrino experiments at the FPF would consist of upgrades of the Run 3 experiments with the FASER$\nu$2 and the FAR Advanced SND (AdvSND) detectors, plus a liquid Argon time projection chamber called FLArE. The FPF location is planned for a site at a distance of $\sim 620-685$ m from the ATLAS interaction point, and will have sizes that allow to cover forward neutrino pseudo-rapidities, $\eta_\nu\gtrsim 7$ \cite{Feng:2022inv}. 
Over a baseline of $620-685$ m, while oscillations among active neutrinos are suppressed, oscillations between active and sterile neutrinos with masses of the order of 10s of eV can be investigated \cite{Bai:2020ukz}.
A second AdvSND detector, called NEAR, installed outside the FPF and closer to one of the LHC IP, is planned to cover neutrino rapidities in the range $4< \eta_\nu < 5$ \cite{Feng:2022inv}.

Neutrinos are also interesting from the point of view of astrophysics \cite{Ackermann:2022rqc}. Measurements of the diffuse flux of astrophysical neutrinos have opened a window on high-energy neutrino sources \cite{IceCube:2013cdw,IceCube:2020acn,IceCube:2020wum}.
An important background to the diffuse flux is the flux of neutrinos produced in cosmic ray interactions in the atmosphere. The so-called ``conventional'' atmospheric flux comes from the decays of charged pions ($\pi^\pm$) and kaons ($K^\pm$) produced in these interactions.
As their decay lengths increase with energy, pions and kaons are apt to loose energy through interactions with other particles before they decay \cite{Lipari:1993hd,Gaisser:2016uoy}. The resulting conventional neutrino flux steeply falls as energy increases.
    
A portion of atmospheric neutrinos also comes from prompt heavy-flavour hadron decays, predominantly from charm mesons. While the production cross sections for heavy-flavour hadrons are smaller than for pions and kaons, 
the flux of prompt atmospheric neutrinos has a harder spectrum than the conventional one due to very short lifetime of heavy-flavour hadrons. 
Accordingly, the prompt neutrino flux becomes more important than the conventional neutrino flux at a sufficiently high energy. One feature of both the prompt and conventional components of the atmospheric neutrino flux is that 
the energies of hadrons most important to the atmospheric neutrino flux carry large fractions of the energies of the cosmic rays primaries impinging on the atmosphere and interacting with the atmospheric nuclei.
This is a consequence of the fact that the spectrum of high-energy cosmic rays falls steeply according to a power law $\sim E^{-2.7}-E^{-3}$.
Various theoretical evaluations 
\cite{Enberg:2008te,Garzelli:2015psa,Garzelli:2016xmx,Benzke:2017yjn,Zenaiev:2019ktw,Ostapchenko:2022thy,Gauld:2015kvh,Bhattacharya:2015jpa,Bhattacharya:2016jce,Goncalves:2017lvq,Fedynitch:2018cbl}
indicate that the flux of prompt atmospheric neutrinos overcomes the conventional one above $E_\nu \sim 10^5 - 10^6$~GeV. 
To date, the atmospheric neutrino spectra inferred by experimental observations seem to be consistent with the conventional component, the flux of which is dominant at relatively low energies. 
Although upper limits have been established \cite{IceCube:2020wum, IceCube:2016umi},
a definitive measurement of the prompt atmospheric neutrino flux has not been made yet. 

Theoretical predictions of the prompt atmospheric neutrino fluxes have large uncertainties, mainly due to a poor understanding of heavy flavour production \cite{Enberg:2008te,Garzelli:2015psa,Garzelli:2016xmx,Benzke:2017yjn,Zenaiev:2019ktw,Ostapchenko:2022thy,Gauld:2015kvh,Bhattacharya:2015jpa,Bhattacharya:2016jce,Goncalves:2017lvq,Fedynitch:2018cbl}.
As we will discuss in the following of this work, the production of high-energy neutrinos in the atmosphere has some kinematic overlap with the production of neutrinos in the forward region at the LHC. 
The   $\sqrt{s}= 14 {\ \rm TeV}$ collision energy 
foreseen for the HL-LHC phase, corresponds to a cosmic-ray nucleon energy $E_p \sim 10^8 {\ \rm GeV}$ on a fixed-target (air) nucleon. Through production and decays of hadrons, these 
high-energy cosmic-ray interactions in the atmosphere yield neutrinos in the energy range where the prompt component is the most important part of the atmospheric neutrino flux. 
With the high lu\-mi\-no\-si\-ty upgrade that will be a factor of 10 times the luminosity of Run 3,
the LHC can produce an abundant number of prompt neutrino events in the forward region, through the interaction of forward prompt neutrinos with the nuclear targets of forthcoming experiments~\cite{Feng:2022inv}. 
Measurements of heavy flavour and/or prompt neutrino production at forward LHC experiments can contribute to reduce the uncertainty in the theoretical predictions of the prompt atmospheric neutrino fluxes.    

The aim of our work is to explore the connections between measurements of production of neutrinos and heavy hadrons at colliders and predictions of prompt neutrino fluxes in the atmosphere, and as a consequence, the potential utility of these measurements for neutrino astronomy applications with particular emphasis on neutrino measurements at the FPF.
Towards this purpose, we examine in detail the overlap of kinematic regimes, extending our results of ref.~\cite{Jeong:2021vqp}. 
Our focus is not on precision predictions of the prompt atmospheric neutrino flux, the topic of many papers on this subject (see, e.g., refs. \cite{Garzelli:2015psa,Bhattacharya:2016jce,Gauld:2015kvh, Garzelli:2016xmx,
Bhattacharya:2015jpa,Enberg:2008te,Benzke:2017yjn,Goncalves:2017lvq,Zenaiev:2019ktw,Ostapchenko:2022thy,Fedynitch:2018cbl}). Instead we aim to demonstrate the extent to which measurements of neutrino fluxes in the very forward region at the FPF could have implications for theoretical predictions of the prompt atmospheric neutrino flux.

We use as a basis for our results a QCD evaluation of charm meson production cross sections, including charm quark production at next-to-leading order (NLO) in perturbative QCD (pQCD) 
\cite{Nason:1989zy,Mangano:1991jk}. We employ as input for this calculation the PROSA parton distribution functions (PDFs) \cite{Zenaiev:2019ktw} with parameters that make the results well-matched to the measurements by the LHCb experiment \cite{LHCb:2013xam,LHCb:2015swx,Aaij:2018afd}. This is not the only PDF set that incorporates LHCb charm data in the fit. Refs. \cite{Gauld:2015yia,Gauld:2016kpd,Bertone:2018dse}  describe the inclusion of these data in two NNPDF sets. This has finally led to gluon PDFs consistent with the PROSA gluon PDF within uncertainties. We use phenomenological fragmentation functions to describe the parton-to-meson transition, and we use analytical formulas for the evaluation of neutrino production in charmed meson decay \cite{Bai:2018xum,Bai:2020ukz}. 
We examine the role played by collisions at different center-of-mass energies $\sqrt{s}$, up to a value of $\sqrt{s} = 100$~GeV foreseen for hadron-hadron interactions at a 
Future Circular Collider (FCC) \cite{FCC:2018vvp,Benedikt:2022wvj}. 
Since the neutrino pseudo-rapidity $\eta_\nu$ is correlated with the charm hadron rapidity $y$ \cite{Bai:2021ira},
we also consider the rapidity $y$ of charm hadrons relevant to prompt atmospheric neutrino flux predictions. In the following, $y$ will be used to refer to the charm hadron rapidity in the proton-proton center-of-mass (CM) frame, or equivalently collider frame.  We will also illustrate the impact of the small-$x$ and large-$x$ PDFs on the prompt atmospheric neutrino fluxes in connection with charm produced at the LHC at different collider rapidities. 

The work is organized as follows. 
In section~2, the semi-analytic approach with $Z$-moments to determine the atmospheric neutrino fluxes from 
a set of 
coupled differential equations called cascade equations is reviewed. The cosmic ray spectra used as input in this work are also discussed in this section. 
In section~3, results for the $Z$-moments for charm hadron production and the atmospheric neutrino fluxes for different regions of $\sqrt{s}$ and different intervals of collider-frame rapidity $y$ of the parent charm hadrons are shown. Parton distribution function (PDF) uncertainties are illustrated and discussed in section~4. Finally, conclusions are drawn in section~5.

\section{Prompt atmospheric neutrino fluxes} 

\subsection{Cascade equations}
The atmospheric neutrino fluxes can be evaluated using the cascade equations and the so-called $Z$-moment method \cite{Lipari:1993hd}. 
The cascade equations account for the propagation of the particles in the atmosphere, and the general expression is given by 
\begin{eqnarray}
\label{eq:cascade}
\frac{d\phi_j(E,X)}{dX}&=&-\frac{\phi_j(E,X)}{\lambda^{\rm int}_j(E)} - \frac{\phi_j(E,X)}{\lambda^{\rm dec}_j(E)}
+ \sum_k S(k\to j)\, , \\   \nonumber
S(k\to j) &=& \int_E^{\infty}dE ' \frac{\phi_k(E',X)}{\lambda_k(E')}
\frac{dn(k\to j;E',E)}{dE}  \, .
\end{eqnarray}
Here, $\phi_j(E,X)$ is the flux of particle $j$ at the column depth $X$, and $\lambda_{j /k}^{\rm int}$ and $\lambda_{j /k}^{\rm dec}$ are the interaction and decay length for particle $j$ (or $k$), respectively.
In the source term $S(k\to j)$, $dn(k \to j;E',E)/dE$ depends on whether $k\to j$ proceeds through interactions or through decays with 
\begin{align}
\label{eq:dnde}
\frac{dn(k\to j;E',E)}{dE} =
\begin{cases}
\frac{1}{\sigma_{kA}(E')}\frac{d\sigma(kA\to jY;E',E)}{dE }
\quad {\rm (interaction)\, }\\
\frac{1}{\Gamma_k(E')}\frac{d\Gamma (k\to jY;E',E)}{dE}\quad {\rm (decay)\,,} 
\end{cases}
\end{align}
therefore
$\lambda_k$ can be an interaction or decay length in the expression for $S(k\to j)$.

It is convenient to introduce the quantity $Z_{kj}(E)$.
This so-called $Z$-moment is related to the spectrum-weighted differential cross section for particle production through interaction or decay and  is defined as 
\begin{eqnarray}
%%%S(k\to j) &\simeq & Z_{kj}(E)\frac{\phi_k(E,X)}{\lambda_k(E)}\, ,\\
Z_{kj}(E) &\equiv& \int _E^{\infty}dE ' \frac{\phi_k(E',0)}{\phi_k(E,0)}
\frac{\lambda_k(E)}{\lambda_k(E')}
\frac{dn(k\to j;E',E)}{dE} \, .
\label{eq:Zmoment}
\end{eqnarray}
Then, the source term $S(k\to j)$ in eq. (\ref{eq:cascade}) can be approximated in terms of the only energy-dependent $Z$-moment, the flux and the interaction (or decay) length of particle $k$ 
according to
\begin{eqnarray}
\label{eq:skj}
S(k\to j) &\simeq & Z_{kj}(E)\frac{\phi_k(E,X)}{\lambda_k(E)}
\end{eqnarray}
under the assumption that $\phi_k (E',X)/\phi_k(E,X)\simeq \phi_k(E',0)/\phi_k(E,0)$. For cosmic-ray nucleons, labeled here with $k=p$, this means that 
$\phi_p (E,X)\simeq \phi_p^0(E) \, f(X)$ where $\phi_p^0(E)=\phi_p(E,0)$ is the cosmic-ray flux at the top of the atmosphere and  $f(X)$ describes its attenuation through the atmosphere which is largely energy independent and only depends on column depth $X$.
The column depth depends on distance from the ground $\ell$ and zenith angle $\theta$ and  is given by $X (\ell, \theta) = \int^\infty_\ell d\ell' \rho (h (\ell', \theta))$, where $h (\ell', \theta)$ is the altitude for a given $(\ell,\theta)$.
The atmospheric density is approximated by $\rho (h) = \rho_0 \exp (-h/h_0)$, where $h_0 = 6.4 {\,\, \rm km}$ and $\rho_0 h_0 = 1300 {\,\, \rm g/cm^2}$.

The coupled cascade equations then consist of equations for incident cosmic rays, produced hadrons and neutrinos from the decays of hadrons. 
The atmospheric neutrino fluxes can be obtained by solving this system of  
coupled cascade equations.  Two approximate solutions in terms of the $Z$-moments and incident cosmic ray spectrum in the low- and high-energy limits can be obtained at the surface of the Earth, with
\begin{eqnarray}
\label{eq:phinu_lo}
\phi_{h\to \nu}^{\rm low} &= &\sum_h \frac{Z_{ph}Z_{h\nu}}{1-Z_{pp}}\phi_p^0\, ,\\ \label{eq:phinu_hi}
\phi_{h\to \nu}^{\rm high} &= &\sum_h \frac{Z_{ph}Z_{h\nu}}{1-Z_{pp}}\frac{\ln(\Lambda_h/\Lambda_p)}
{1-\Lambda_p/\Lambda_h}\frac{\epsilon_h}{E}
\phi_p^0 \, ,
\end{eqnarray}
given the effective interaction length $\Lambda_k = \lambda_k^{\rm int} / (1-Z_{kk})$.  
The proton--proton and proton--deuterium total inclusive cross sections per nucleon are quite similar to each other. 
Charm hadron production is largely independent of whether the interacting particles are protons or neutrons. For these reasons, we have denoted the cosmic ray nucleons with $p$ in the $Z$-moments to evaluate the prompt atmospheric lepton flux. In practice, cosmic rays have several mass components, as discussed below in section \ref{subsec:crspectrum}. The cosmic ray flux at the top of the atmosphere as a function of energy per nucleon  $\phi_p^0$ depends on the cosmic ray mass composition.

The high-energy and low-energy regimes are 
separated by 
$\epsilon_k\simeq(m_k c^2h_0/c\tau_k)$, the critical energy for each hadron $k$. The hadron $k$ decays dominantly in the low energy regime ($E \ll \epsilon_k$), while it tends to interact at energies higher than $\epsilon_k$.
In the end, the resulting neutrino fluxes for each neutrino flavour can be obtained by
\begin{equation}
\phi_\nu = \sum_h \frac{\phi_{h\to \nu}^{\rm low}\phi_{h\to \nu}^{\rm high}}{(
\phi_{h\to \nu}^{\rm low}+\phi_{h\to \nu}^{\rm high})}\ .
\label{eq:phinu}
\end{equation} 

For the ($\nu_e$ + $\bar{\nu}_e$) and ($\nu_\mu$ + $\bar{\nu}_\mu$) prompt neutrino fluxes, the contributed hadrons $h$ are heavy-flavour hadrons, predominantly charmed hadrons, i.e., $h_c=D_0$, $D^+$, $D_s^+$ and $\Lambda_c^+$ and their antiparticles. 
Their critical energies are in the range of $\epsilon_{h_c}\sim 3.7\times 10^7- 2.6\times 10^8$ GeV.
Most of the neutrino energies of interest here, in particular $E_\nu<10^7$ GeV, are below $\sim\epsilon_{h_c}$, where eq. (\ref{eq:phinu_lo}) applies, so the approximation of eq. (\ref{eq:phinu}) does not impact our conclusions. The prompt 
neutrino fluxes described by  eq. (\ref{eq:phinu_lo}) are isotropic.

While the $Z$-moment method relies on approximations as discussed above including those in eqs. (\ref{eq:skj}-\ref{eq:phinu_hi}), it gives results in agreement with those from more direct methods of determining the atmospheric lepton fluxes. A comparison of atmospheric lepton fluxes from pions and kaons using the $Z$-moment method \cite{Gaisser:2019xlw} with the atmospheric lepton fluxes determined from the step-wise solution of the Matrix Cascade Equation (\textsc{MCEq}) method \cite{Fedynitch:2018cbl} shows agreement of these predictions. They agree to within 5-20\%, depending on the zenith angle \cite{Gaisser:2019xlw}. The atmospheric flux fractions of muons and muon neutrinos from charm decays as a function of energy show a very good agreement between the $Z$-moment method and \textsc{MCEq} \cite{Gaisser:2019xlw}.
In ref. \cite{Gondolo:1995fq}, the atmospheric lepton fluxes from the $Z$-moment method and from a Monte Carlo simulation of the showers that also included charm contributions are shown to be in good agreement, to within $\sim 20\%$.  Since we focus on the relevance of different kinematic regions for charm production and decay at colliders as well as their impact on the prompt atmospheric neutrino flux, the $Z$-moment approximation is sufficient. Indeed, the $Z$-moment approximation is used in many predictions of the prompt atmospheric neutrino flux
\cite{Garzelli:2015psa,Bhattacharya:2016jce,Gauld:2015kvh, Garzelli:2016xmx,
Bhattacharya:2015jpa,Enberg:2008te,Benzke:2017yjn,Goncalves:2017lvq,Zenaiev:2019ktw,Ostapchenko:2022thy}. New measurements of forward charm production may motivate a more detailed accounting of the uncertainties in the particle physics inputs and of those associated with the approximate solutions.

\subsection{Cosmic ray spectrum}
\label{subsec:crspectrum}

For the cosmic-ray all-nucleon fluxes as a function of the energy per nucleon $E$, several parameterizations exist. A parameterization often used for comparisons is a broken power law (BPL) spectrum, 
\begin{align}
\label{eq:cr}
\phi^0_p(E) \, 
[\text{cm}^{-2} \, \text{s}^{-1} \, \text{sr}^{-1} \, (\gev/A)^{-1} ] 
= 
\begin{cases}
1.7 \, E^{-2.7}  \quad  &\text{for } E<5 \cdot 10^6 \,\, \gev \\ 
174 \, E^{-3}           &\text{for } E>5 \cdot 10^6 \,\, \gev\,. 
\end{cases}
\end{align}
While the BPL approximates the all-particle cosmic ray spectrum, it also approximates 
the all-nucleon energy spectrum only if the cosmic-ray composition consists of only nucleons (or protons). Recent parameterizations of the cosmic-ray nucleon spectrum take into account different sources and compositions. Currently, the ultra-high-energy composition of cosmic rays is not clearly identified (see, e.g., ref. \cite{Coleman:2022abf} for a review). 
Although the origin of cosmic rays, together with the composition, is still a topic of intense debate, two models frequently used, referred to as H3p and H3a, involve three possible cosmic-ray source components: supernova remnants, other galactic sources and extra-galactic sources \cite{Gaisser:2011klf}. 
The difference between the two spectrum models is the component of high-energy cosmic rays originated from the extra-galactic sources, only protons for H3p and a mixed composition involving heavier nuclei for H3a. 
It is also worth mentioning that the most recent estimates from cosmic ray extended air shower experiments, such as the Pierre Auger Observatory \cite{Kampert:2012mx} and Telescope Array \cite{TelescopeArray:2020bfv}, point towards a mixed composition at the highest energies (see also refs. \cite{Coleman:2022abf,Bowman:2022okd}). 

Although the BPL is based on over-simplified assumptions, it is
commonly used for comparisons with other existing evaluations. In this work, 
we use the BPL as the reference cosmic ray nucleon spectrum to evaluate and illustrate which ranges of $\sqrt{s}$ of cosmic ray nucleon interactions with air nucleons and which ranges of charmed-meson hadron-collider (i.e. nucleon-nucleon center-of-mass) rapidities play the most relevant role in determining the prompt atmospheric neutrino fluxes at different neutrino energies. 
We also evaluate the prompt neutrino fluxes with the H3p and H3a spectra and compare with the predictions obtained using the BPL spectrum. 

\section{{\it Z}-moments and atmospheric neutrino fluxes from charm}
\subsection{{\it Z}-moments for charm hadron production}

As discussed in  the previous section,
atmospheric neutrino fluxes can be obtained in terms of $Z$-moments: $Z_{pp}$ for proton regeneration, $Z_{ph}$ and $Z_{hh}$ for hadron production and regeneration, and $Z_{h\nu}$ for decays to neutrinos (see eqs. (\ref{eq:phinu_lo}) and (\ref{eq:phinu_hi})). 
We use standard inputs for $Z_{pp}$ and $Z_{hh}$, as described in ref. \cite{Bhattacharya:2015jpa}. An energy independent scaling value of $Z_{pp}=0.263$ \cite{Gaisser:2016uoy} is modified by using a weakly energy dependent $pA$ cross section and energy independent distribution in $x_E$, the ratio of the energy of the leading outgoing proton to the incoming proton.
This yields a range of $Z_{pp}$ values, from $Z_{pp}(10^3\ {\rm GeV})=0.271$ to $Z_{pp}(10^8\ {\rm GeV})=0.231$. For the energy range of interest, the uncertainty in $Z_{pp}$ translates to an overall factor since $\phi_{h\to \nu}^{\rm low}$ is proportional to $(1-Z_{pp})^{-1}$.
With a similar approximation for $Z_{hh}$ for $D$-mesons approximated by $Z_{KK}$, the scaling value of $Z_{KK}=0.211\simeq Z_{DD}$ \cite{Gaisser:2016uoy} becomes $Z_{DD}(10^3\ {\rm GeV})=0.217$ and increases with energy to $Z_{DD}(10^8\ {\rm GeV})=0.176$.
While the evaluation of $Z_{h\nu}$ is relatively straightforward, there are large uncertainties in the $Z$-moments for the production of heavy-flavour hadrons ($Z_{ph}$) 
which have a stronger energy dependence than $Z_{pp}$.
The uncertainties in the $Z$-moments 
translate into uncertainties in the predictions of prompt neutrino fluxes. We note that there is work on using data-driven models for hadronic interactions of protons, neutrons, pions and kaons \cite{Fedynitch:2022vty} that can inform assessments of the uncertainty in $Z_{pp}$. The uncertainties in $Z_{pp}$ and $Z_{hh}$, important for the overall normalization of the prompt atmospheric neutrino flux over the full energy range, do not change our conclusions about the kinematic regions that contribute to this flux. 

The uncertainty in the theoretical predictions of $Z_{ph}$ mainly come from the truncation of the QCD perturbative expansion for differential cross sections of heavy-flavour production at next-to-leading order (NLO) \cite{Nason:1989zy,Mangano:1991jk}. The QCD scale dependence through the renormalization scale ($\mu_R$) and factorization scale ($\mu_F$) is particularly large for charm production, 
also due to the smallness of the charm quark mass. The value of the latter is well above the $\Lambda_{QCD}$ value to allow for the application of a perturbative treatment, but still close enough to it to lead to relatively large $\alpha_s(\mu_R \sim m_c)$ values and to the evaluation of parton distribution functions at characteristic scales $\mu_F \sim m_c$  
close to the lowest extremes of their characteristic range of evolution. The QCD scale dependence primarily changes the normalization, but not the shape of the resulting energy distributions of forward neutrinos from charm \cite{Bai:2021ira}.
We mitigate this uncertainty by anchoring our predictions 
to LHCb data on open charm production for collider rapidities of charm hadrons in the range $2\leq y\leq 4.5$ \cite{LHCb:2013xam,LHCb:2015swx,Aaij:2018afd}. We evaluate charm production in the hadron-collider frame (nucleon-nucleon center-of-mass frame CM), then boost to the fixed-target frame to calculate the $Z$-moments. Rapidities change under boosts from collider frame to fixed-target frame by $\Delta y={\rm tanh}^{-1}\beta\simeq\ln(\sqrt{s}/m_p)$,
so that the CM charm hadron $y$ can be easily converted to the fixed-target frame. Since we focus here on connections to LHC experiments, we only reference CM rapidities in case of charm hadrons. 

In our previous works for estimating the prompt neutrino fluxes generated at the LHC~\cite{Bai:2020ukz, Bai:2021ira,Bai:2022jcs}, we evaluated the differential cross sections for charm meson production including NLO QCD corrections in the collinear approximation in a 3-flavour number scheme framework and compared with the LHCb data. % at 7 TeV and 13 TeV. 
In ref. \cite{Bai:2021ira}, in particular, we used the PROSA 2019 PDFs fitted in this same fixed-flavour-number scheme (FFNS)~\cite{Zenaiev:2019ktw} and found that the QCD scales of ($\mu_R$, $\mu_F$) = (1, 2)~$m_T$ with the transverse mass $m_T = (p_T^2 + m_c^2)^{1/2}$, and the intrinsic transverse momentum smearing $\langle k_T \rangle = 1.2~\gev$ yielded results that are reasonably well-matched to the LHCb data. These results for LHCb \cite{Bai:2021ira} are consistent with results using POWHEG \cite{Frixione:2007vw} plus PYTHIA \cite{Sjostrand:2019zhc}. The strategy \cite{Bai:2020ukz, Bai:2021ira,Bai:2022jcs} to use the most forward charm production data available at the LHC to anchor the QCD prediction for $y=2.0-4.5$ is also used as a starting point by other authors using other approaches (e.g., $k_T$ factorization \cite{Bhattacharya:2023zei}, intrinsic charm and recombination mechanisms  \cite{Maciula:2022lzk} and in refs. \cite{Anchordoqui:2021ghd,Feng:2022inv}) 
%%%to extrapolate to larger charm rapidities 
to make predictions for charm and prompt neutrino flux at rapidities larger
than the LHCb ones. 
In this work, we use NLO QCD in the collinear factorization approximation, using the aforementioned input parameter set, plus the PROSA 2019 PDFs as default inputs in evaluating theoretical predictions of the prompt atmospheric neutrino flux. While not guaranteed to work for extremely high $y$, the agreement of our evaluations with LHCb charm hadron $p_T$ distributions from $y=2.0-2.5$ to $y=4.0-4.5$ gives some confidence that the approach leads to reliable results even for $y>4.5$. In fact, we show below that the most important charm hadron rapidity region for the prompt atmospheric neutrino flux is just adjacent to the LHCb measurements, namely, $4.5 < y < 7.2$ instead of involving extremely forward rapidities, for prompt atmospheric neutrinos in the energy range of interest explored by current neutrino telescopes.

\begin{figure}
    \centering
       \includegraphics[width=.65\textwidth]{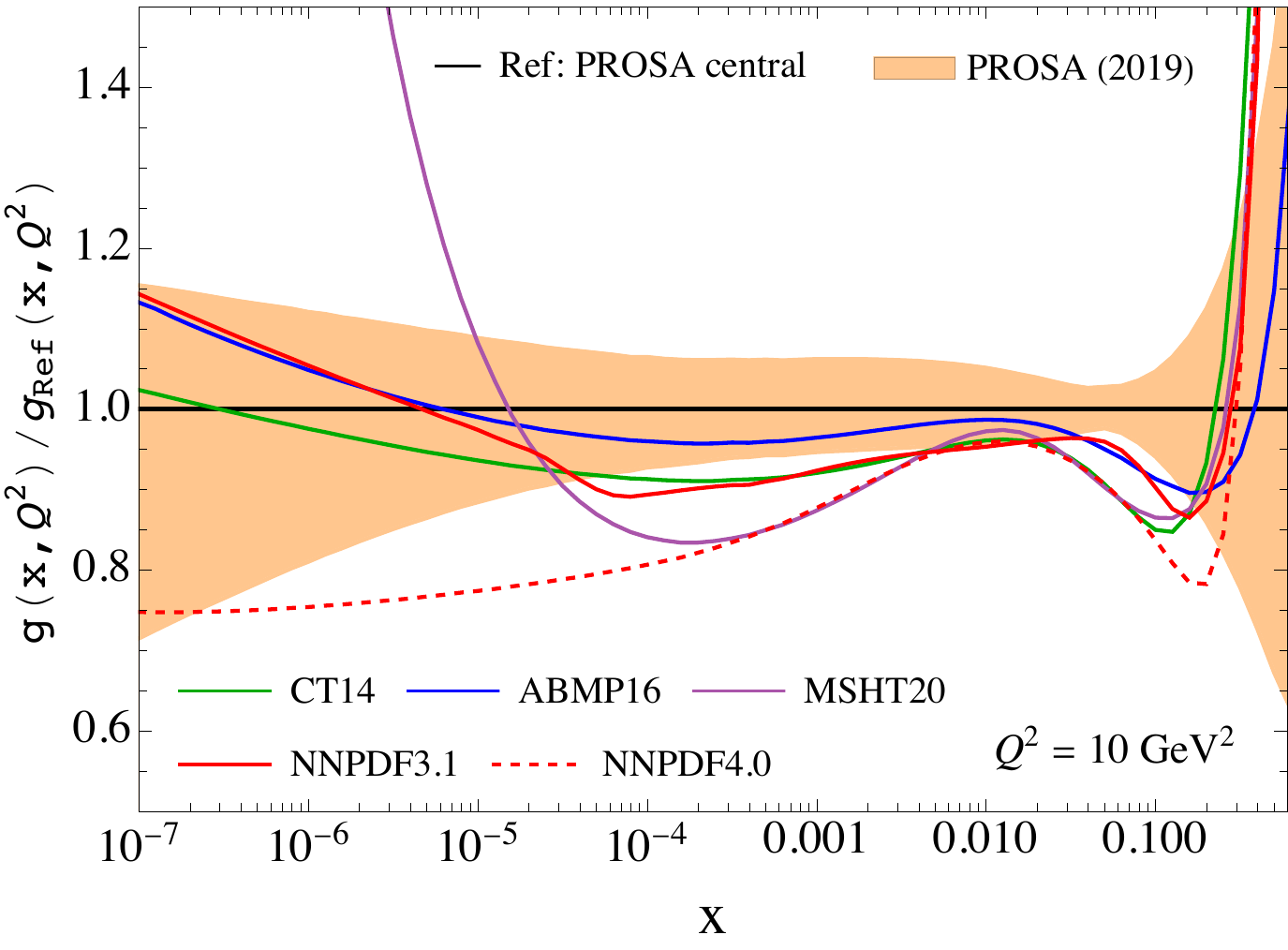}
       \caption{
       Gluon distribution function uncertainty for $Q^2=10$ GeV$^2$ evaluated using the 40 different eigenvectors in the PROSA PDF set (orange band), 
       normalized to the best fit of the PROSA FFNS (2019) PDF set \cite{Zenaiev:2019ktw}.
The uncertainty band is determined from the appropriate combination of fit, model and parameterization uncertainties following the PROSA prescription (see, e.g., Appendix A of ref. \cite{Bai:2021ira}). The ratios of the central NLO gluon distributions in the same flavour number scheme by 
       CT14~\cite{Dulat:2015mca} (green), ABMP16~\cite{Alekhin:2018pai} (blue), MSHT20~\cite{Cridge:2021qfd} (purple), NNPDF3.1~\cite{NNPDF:2017mvq} (red) and NNPDF4.0~\cite{NNPDF:2021njg} (red dashed) to the PROSA best fit are also shown. 
        } 
       \label{fig:pdf-range}
     \end{figure}

Most of the contributions to the cross sections for charmed hadron production in nucleon-nucleon collisions at high energies come from gluon interactions.
Figure~\ref{fig:pdf-range} shows the ratio of the gluon distribution functions $x g (x, Q^2)$ to the same quantity using the best fit of the PROSA PDF set for $Q^2 = 10~\gev^2$. 
The uncertainty from the 40 variations within the PROSA PDF set is presented with the orange band, and the central fits of other 3-flavour NLO PDFs, i.e.  CT14~\cite{Dulat:2015mca}, ABMP16~\cite{Alekhin:2018pai}, MSHT20~\cite{Cridge:2021qfd}, NNPDF3.1~\cite{NNPDF:2017mvq} and NNPDF4.0~\cite{NNPDF:2021njg}, are also shown for comparison. 
The experimental data used in the determination of existing PDFs cover a limited $x$ region, mainly $10^{-4} \lesssim x \lesssim 10^{-1}$ from $ep$ deep inelastic scattering (DIS) at HERA and $x \gtrsim 10^{-5}$ from $pp$ scattering at the LHC \cite{Workman:2022ynf}.
By using LHC heavy-flavor production data \cite{LHCb:2013xam,LHCb:2013vjr,LHCb:2015swx,LHCb:2016ikn,ALICE:2017olh,ALICE:2019nxm}, PDF fits can be extended to $x\sim 3\times 10^{-6}$.
Present constraints for $x \gtrsim 0.1$ come from LHC data and old data from fixed-target experiments involving nuclear targets. 
Large uncertainties remain, particularly for $x\gtrsim 0.6$.
Due to combination of a lack of coverage of the $x$ range and insufficient data, the PDFs are currently not well constrained for both the very low-$x$ and very large-$x$ regions, which leads to the large uncertainty band as indicated in the figure.

The NNPDF4.0 NLO set is an outlier in the comparison of Figure~\ref{fig:pdf-range}. This new set differs from the NNPDF3.1 in its machine learning methodology and the addition of 44 new data sets \cite{NNPDF:2021njg}. The data sets on inclusive jet, dijet and $t\bar{t}$ production play a role in determining the gluon distribution. These processes are better described by NNLO calculations than by NLO calculations, so the NLO fit to these data is less reliable than the NNLO fit.

Another PDF outlier set is the MSHT20 set \cite{Cridge:2021qfd}.  This fit uses a very flexible parameterization of the PDFs and does not include the LHCb open heavy-flavor production data that constrain $x\lesssim 10^{-5}$. The large departure of the gluon PDF from the other sets for $x\lesssim 10^{-5}$ visible in Figure~\ref{fig:pdf-range} likely arises from this flexible parameterization. 

\begin{figure}
 \centering      
       \includegraphics[width=.49\textwidth]{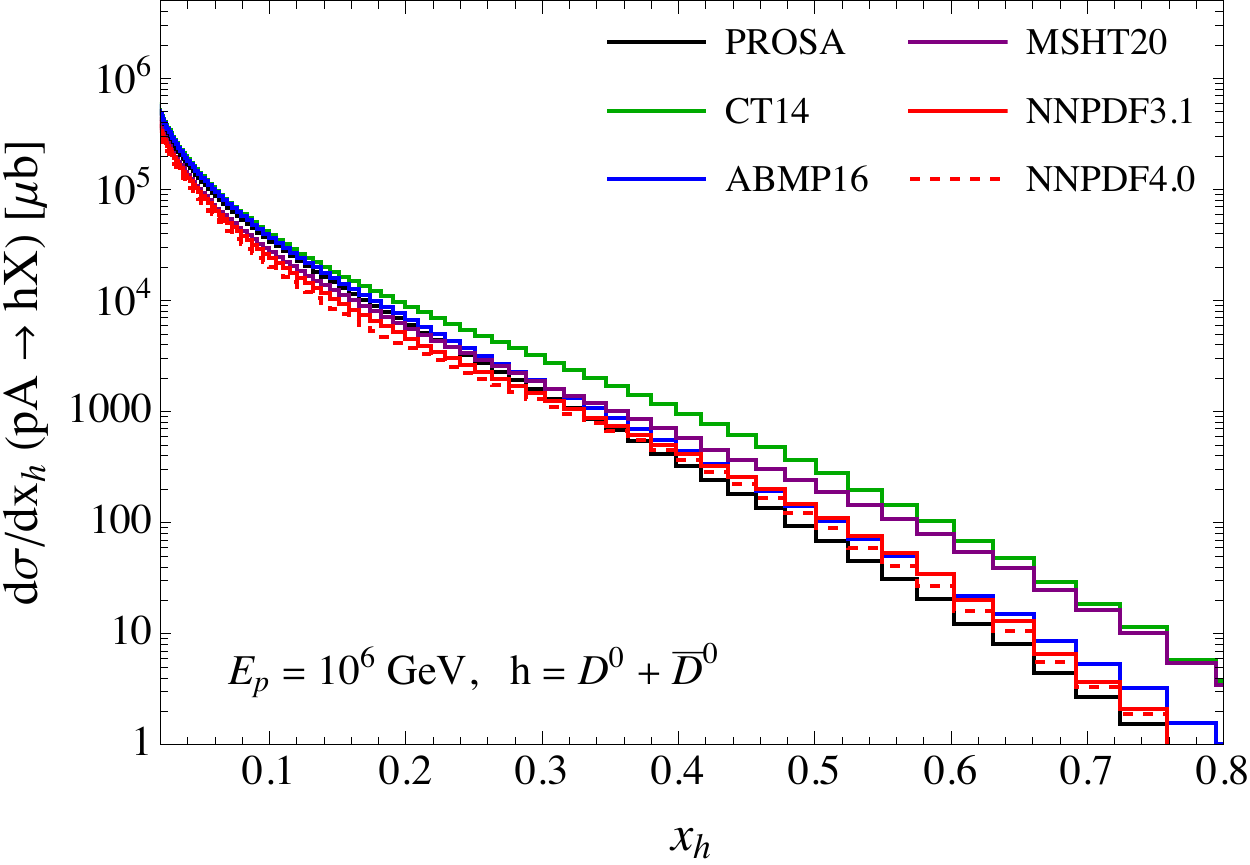}
       \includegraphics[width=.49\textwidth]{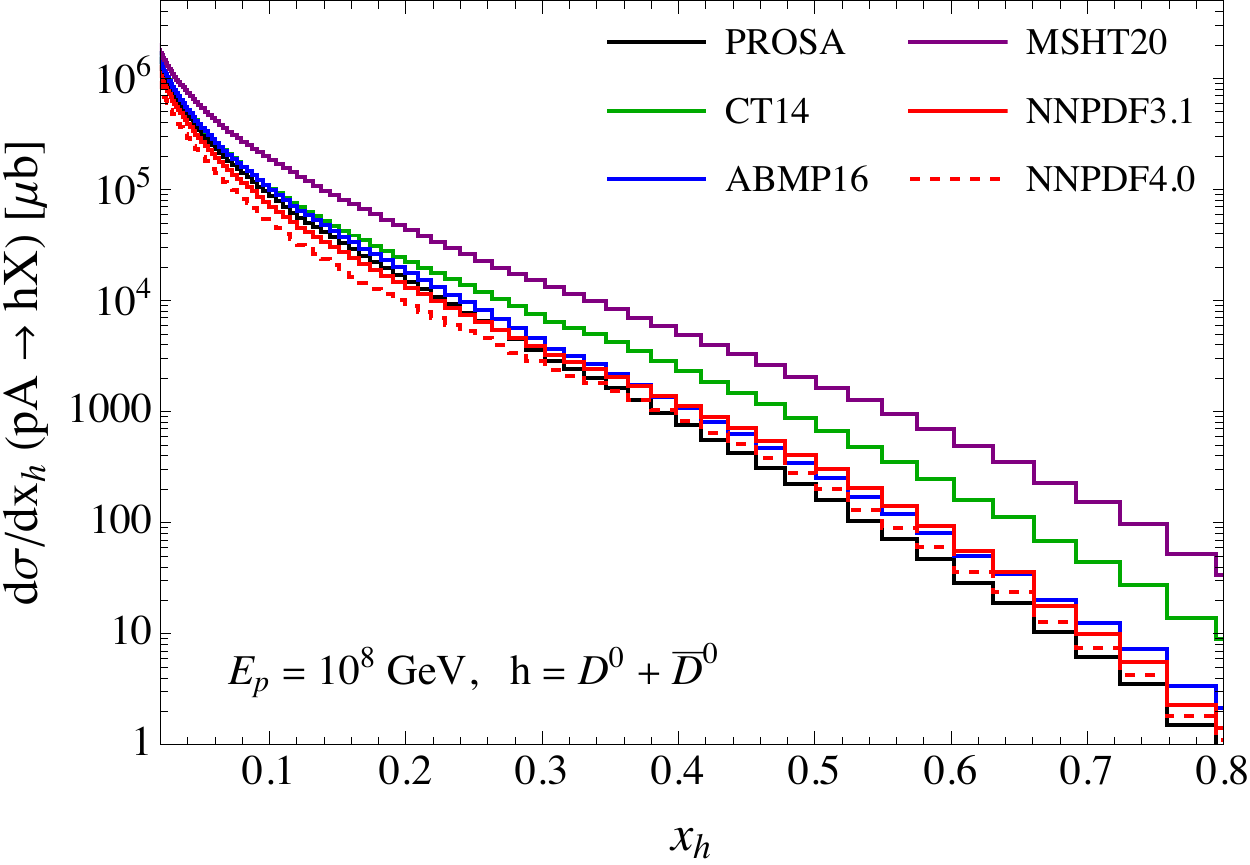}       
       \caption{The differential cross section for the sum of $D^0$ plus $\bar{D}^0$ production in $p$-Air collision, $d\sigma / dx_h$ as a function of $x_h=E_h/E_p$ for $E_p = 10^6 \ \gev$ and $10^8 \ \gev$ evaluated with the central 3-flavour NLO PDF sets from the PROSA \cite{Zenaiev:2019ktw}, CT14 \cite{Dulat:2015mca}, ABMP16 \cite{Alekhin:2018pai},  MSHT20~\cite{Cridge:2021qfd}, NNPDF3.1 \cite{NNPDF:2017mvq} and NNPDF4.0 \cite{NNPDF:2021njg} groups.
} 
       \label{fig:dsdxf}
     \end{figure}

Figure~\ref{fig:dsdxf} shows the differential cross sections for charm meson production $d\sigma / d x_h$ evaluated using the best fits of the different PDFs shown in figure~\ref{fig:pdf-range}.
We present the predictions for the $D^0$ meson for $E_p = 10^{6} {\, \rm GeV}$ and $10^{8} {\, \rm GeV}$ as representatives, and multiply by a factor of~2 to approximately account for both $D^0$ and $\bar{D}^0$ production neglecting charge asymmetries in the production of heavy quarks and in their fragmentation. We implemented the latter through Peterson fragmentation functions \cite{Andersson:1978vj}, identical for the $c$ and $\bar{c}$ quarks.
However, we note that there are indications of charmed meson/antimeson production asymmetries at LHCb \cite{LHCb:2012swq,Aaij:2018afd}, to be better investigated with forthcoming higher-statistics data. 
In the evaluation, to approximate the production in $p$-Air collisions, the cross sections for hadron production in $pp$ collisions are scaled by the average atomic number $\langle A \rangle =14.5$ of air.
The variable $x_h$ is the energy fraction of the produced hadron, $x_h \equiv E_h/E_p$.

The charm mesons from cosmic ray interactions in the atmosphere are produced in the forward direction. 
In this case, the momentum fractions of the partons involved in the interactions from the two sides have typically very different values: the parton momentum fraction from the incident cosmic ray $(x_1)$ is large, whereas the one from the target nucleon in the air nucleus  ($x_2$) is very small.
More specifically, while the former ($x_1$) is approximated to $x_h$, the latter ($x_2$) can be as small as ${\mathcal O} (10^{-6})$ when $E_p = 10^8 \ \gev$ and even smaller for higher incident cosmic-ray nucleon energies. 
As discussed above and shown in figure~\ref{fig:pdf-range}, in such low-$x$ regions as well as for $x \gtrsim 0.1$, the PDFs are not well constrained by the experimental data. 
This implies that the difference between the predictions evaluated with the different central PDF sets becomes larger as $x_h$ increases, due to the combined effects of low-$x$ and large-$x$ PDF behaviour, as shown in figure~\ref{fig:dsdxf}.

\begin{figure}
 \centering
       \includegraphics[width=.7\textwidth]{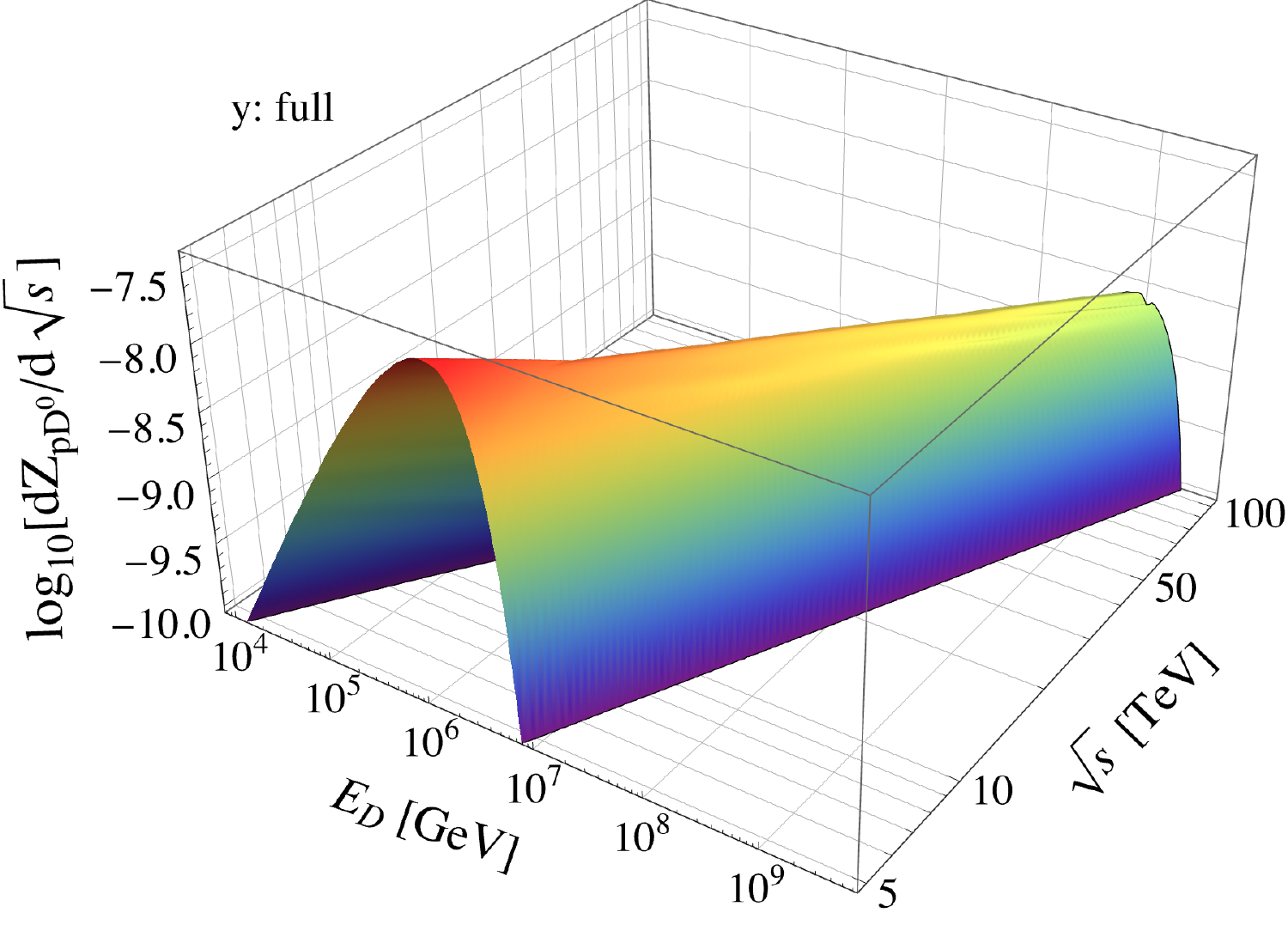}  
       \includegraphics[width=.11\textwidth]{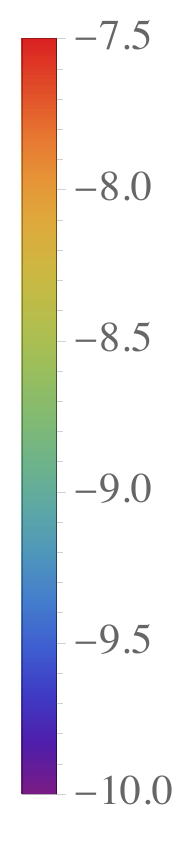}        
       \caption{
          Distribution of differential $Z$-moment for $D^0+\bar{D}^0$ production in proton-air collisions as a function of both the center-of-mass collision energy $\sqrt{s}$ and the $D^0\ (\bar{D}^0)$ meson energy, ${\rm d}Z_{pD^0}(E_D,\sqrt{s})/{\rm d}\sqrt{s}$, plotted on $\log_{10}$ scales. The BPL spectrum is used as input for the cosmic ray flux.
        } 
       \label{fig:dZpD0_rsdist}
     \end{figure}

The $Z$-moment for hadron production in eq. (\ref{eq:Zmoment}) and (\ref{eq:skj}) can be expressed in terms of the energy fraction $x_h$ as
\begin{eqnarray}
Z_{ph}(E_h)  =
\int_{0}^{1} \frac{d x_h}{x_h} \frac{\phi_p^0(E_h/x_h)}{\phi_p^0(E_h)}
\frac{1}{\sigma_{pA}(E_h)}
\frac{d \sigma(pA \to hX)}{d x_h}  \ .
\label{eq:Zprod}
\end{eqnarray}
One of goals of this work is to investigate the impact of the range of collider kinematic variables on the predictions of the prompt atmospheric neutrino fluxes. 
As mentioned earlier in the introduction, a center-of-mass (CM) energy of $\sqrt{s} = 14 \ \tev$, foreseen for HL-LHC, corresponds to $E_p \sim 10^8 \ \gev$ in the fixed-target frame. The next-generation collider FCC-hh aims to increase the energy up to $\sqrt{s} = 100 \ \tev$, equivalent to $E_p \sim 5\times 10^9 \ \gev$ in the fixed-target frame.  
The energy range relevant for probing prompt atmospheric neutrinos is $10^5  \lesssim E_\nu / \gev \lesssim 10^{8}$. Neutrinos with energies greater than $10^7$ GeV are typically produced by $pp$ interactions with $\sqrt{s}>14$ TeV, as we will show in the next section.
Measurements of the heavy-flavour production at the HL-LHC and the FCC will provide useful information for better predictions of the prompt atmospheric neutrino fluxes. 
Here, to relate to collider variables in a more straightforward way, we first convert the expression in eq. (\ref{eq:Zprod}) into integration over the CM collision energy $\sqrt{s}$, converting the $x_h=E_h/E_p$ limits of $x_h$ varying between 0 and 1 to  limits on $\sqrt{s}$, varying between $\sqrt{s}_{\rm min} = \sqrt{2\, m_p E_h}$ and $\infty$, by writing 
\begin{equation}
Z_{ph}(E_h)  =
\int_{\sqrt{s}_{\rm min}}^{\infty} 
\frac{d\sqrt{s}}{\sqrt{s}/2} \
\frac{\phi_p^0(s/2 m_p)}{\phi_p^0(E_h)} \
%\frac{1}{\sigma_{pA}(E_h)}
%\frac{2}{\sqrt{s}}
%\nonumber &\times&
\frac{d \sigma(pA \to hX; E_p=\frac{s}{2m_p}, x_h=\frac{2m_p E_h}{s})}{\sigma_{pA}(E_h) \ d x_h}\,. \label{eq:Zprod-rs}
\end{equation}
We show the differential $Z$-moments for $D^0+\bar{D}^0$ production as a function of both the center-of-mass energy $\sqrt{s}$ and the energy of the produced charm mesons $E_D$ in figure~\ref{fig:dZpD0_rsdist}. We present the predictions  for $D^0+\bar{D}^0$ production as representative of other charmed-hadrons for illustration purposes.
In the evaluation, the BPL spectrum in eq. (\ref{eq:cr}) is used as input for incident cosmic ray flux. For a given $E_D$, figure \ref{fig:dZpD0_rsdist} shows the range of $\sqrt{s}$ values that contribute to $Z_{pD^0}(E_D)$, which increases to higher center-of-mass collision energy with increasing $E_D$.

\begin{figure}
 \centering
    \includegraphics[width=.49\textwidth]{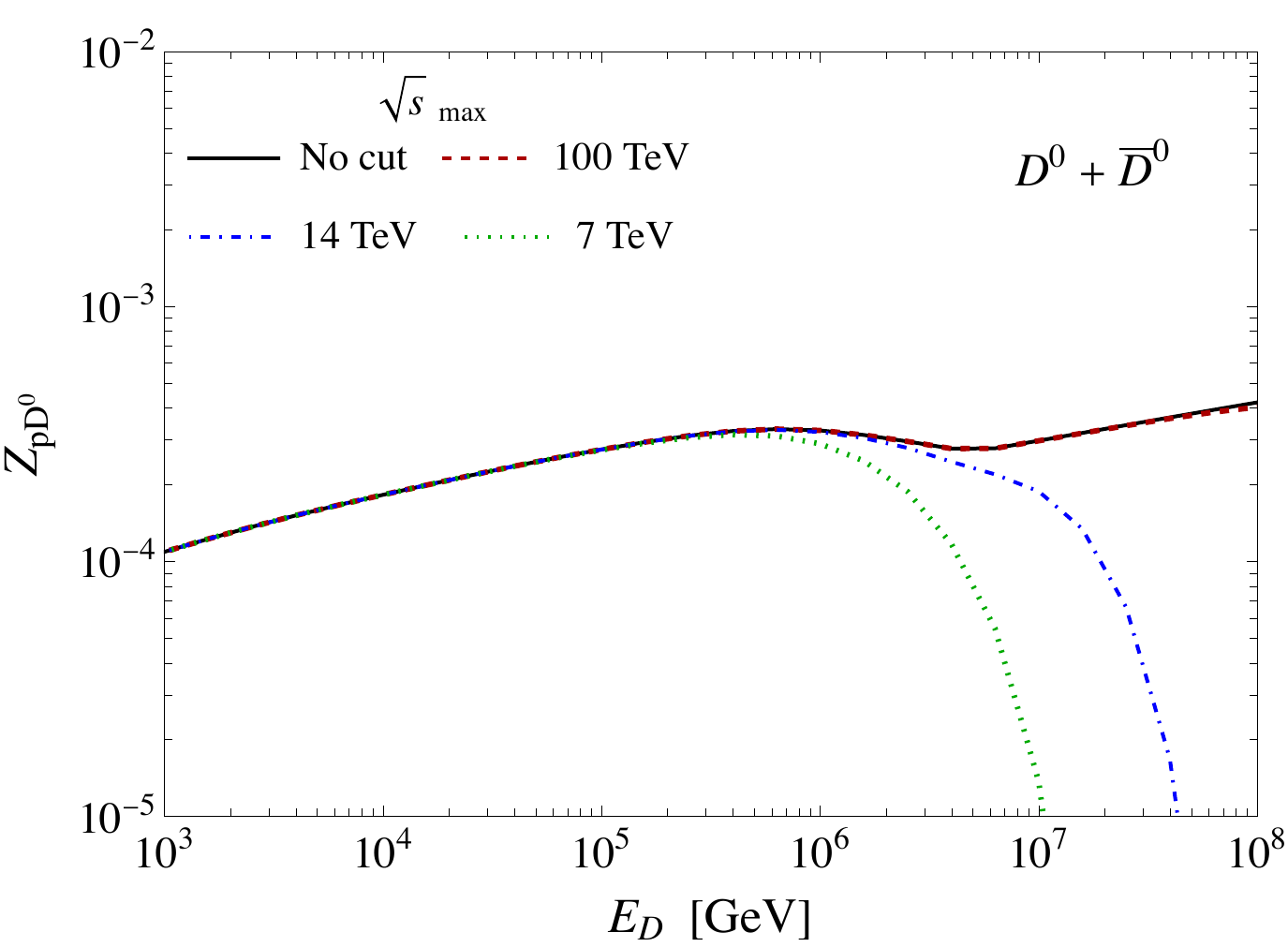}       
    \includegraphics[width=.49\textwidth]{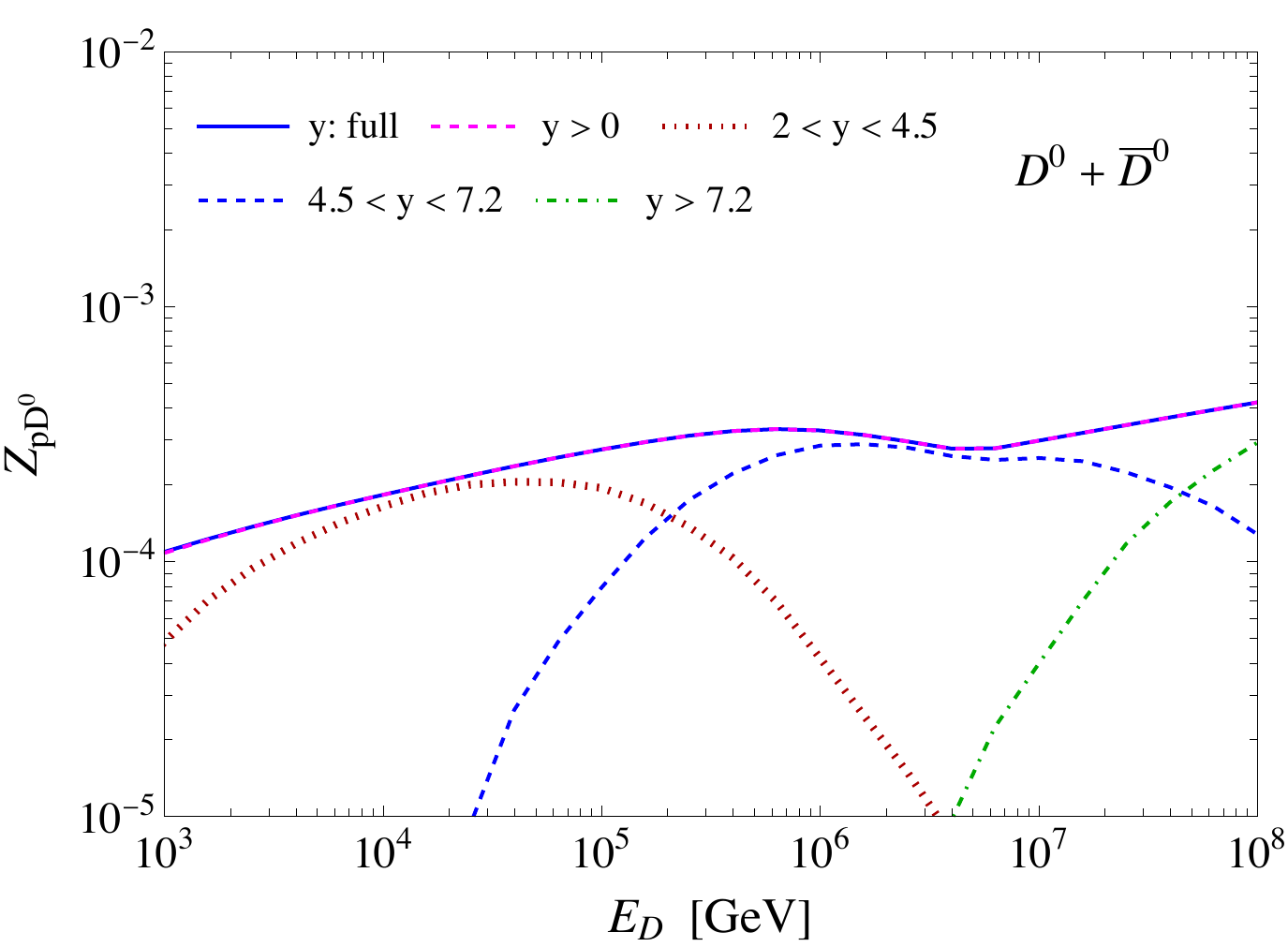}      
   \caption{The production moment $Z_{pD^0}$ for $D^0+\bar{D}^0$ in proton-Air collisions with a broken power law cosmic ray spectrum, evaluated for different values of the maximum hadronic collision energy $\sqrt{s}_{\rm max}$ (left) and  charm hadron CM rapidity ranges (right).  }
\label{fig:ZpD0_rs-Y}
\end{figure}

The left panel of figure \ref{fig:ZpD0_rs-Y} shows the impact of the range of $\sqrt{s}$ in eq. (\ref{eq:Zprod-rs}) on
the $Z$-moments for $D^0 + \bar{D}^0$ meson production as a function of the $D^0$ ($\bar{D}^0$) meson energy, considering the cases of upper extreme of integration fixed to $\sqrt{s}_{\rm max}=$ 7, 14 and 100 TeV, respectively, and for the full range of $\sqrt{s}$. 
The first two cuts on the collision energy are to show the reach in $E_D$ of Run 1 and HL-LHC, respectively (the maximum reach of Run 2 and Run 3, not shown explicitly in the plot, are indeed between these two collision energies). 
The potential reach of the FCC is $\sqrt{s}_{\rm max}=$ 100 TeV.
One can see that the $Z_{pD^0}$-moment evaluated 
by integrating over CM energies up to a maximum value of 14 TeV 
is sizable for $E_D$ smaller than $10^7~\gev$, while the cross sections for charmed-hadron production with center-of-mass energies up to $\sqrt{s}_{\rm max} = 100~\tev$ contribute to most of the $Z$-moments for higher $E_D$. A cut-off at $\sqrt{s}_{\rm max} = 100~\tev$ becomes apparent only for $E_D\gtrsim 10^8$ GeV, consistent with the range of $E_D/E_p$ in the fixed-target frame that dominate the evaluation of $Z_{pD}$ \cite{Goncalves:2017lvq,Pasquali:1998ji}.

Both panels of figure \ref{fig:ZpD0_rs-Y} show a dip in $Z_{pD^0}$ at $E_D\sim$ a few times $10^6$ GeV. The $Z$-moment is a cosmic ray flux-weighted quantity (see eq. (\ref{eq:Zmoment})), so the change in spectral index of the BPL spectrum in eq. (\ref{eq:cr}) is reflected here. A similar effect is seen in the $Z$-moments evaluated with other cosmic ray spectra, as shown in, e.g., ref. \cite{Bhattacharya:2015jpa}.

In the right panel of figure~\ref{fig:ZpD0_rs-Y}, we show the $Z$-moments for the $D^0+\bar{D}^0$ produced in different ranges of CM charm hadron rapidity $y$ using the results integrated over the full range of $\sqrt{s}$. Of course, in 
colliders, charm mesons can be produced in both forward and backward directions. Since we use the collider rapidity, the prediction denoted as `$y$: full' includes 
both positive and negative values of CM charm meson $y$.   
We denote charm hadron rapidity in the fixed-target frame with $y^{\rm fixed}$. 
Charm hadrons with negative $y$ such that $|y|$ is large in the collider frame,
with the appropriate Lorentz boost along the beam axis, translate to low energies in the fixed-target frame since $y^{\rm fixed}\simeq y+\ln(\sqrt{s}/m_p)$, as noted above.
The prediction with collider-frame charm-hadron rapidity $y>0$ is smaller than the prediction without rapidity cuts by $\sim 10 \%$ at $E_D = 10^2 \ \gev$, while these predictions overlap for the whole $E_D$ energy range considered in the figure. 
Thus, for the energies relevant to the prompt atmospheric neutrino fluxes, the $Z$-moments can be described by charm mesons produced in the forward direction in the collider frame ($y>0$). 
We separate further the charm hadron collider-frame rapidity range into three parts, $2 < y < 4.5$, $4.5 < y < 7.2$ and $y > 7.2$.
The range of charm-hadron rapidity $2 < y < 4.5 $ is covered by the LHCb experiment.
This range is the most forward region for heavy-flavour production probed to date at the LHC.

\subsection{$Z$-moments and FPF neutrinos}

\begin{figure}
 \centering
       \includegraphics[width=.49\textwidth]{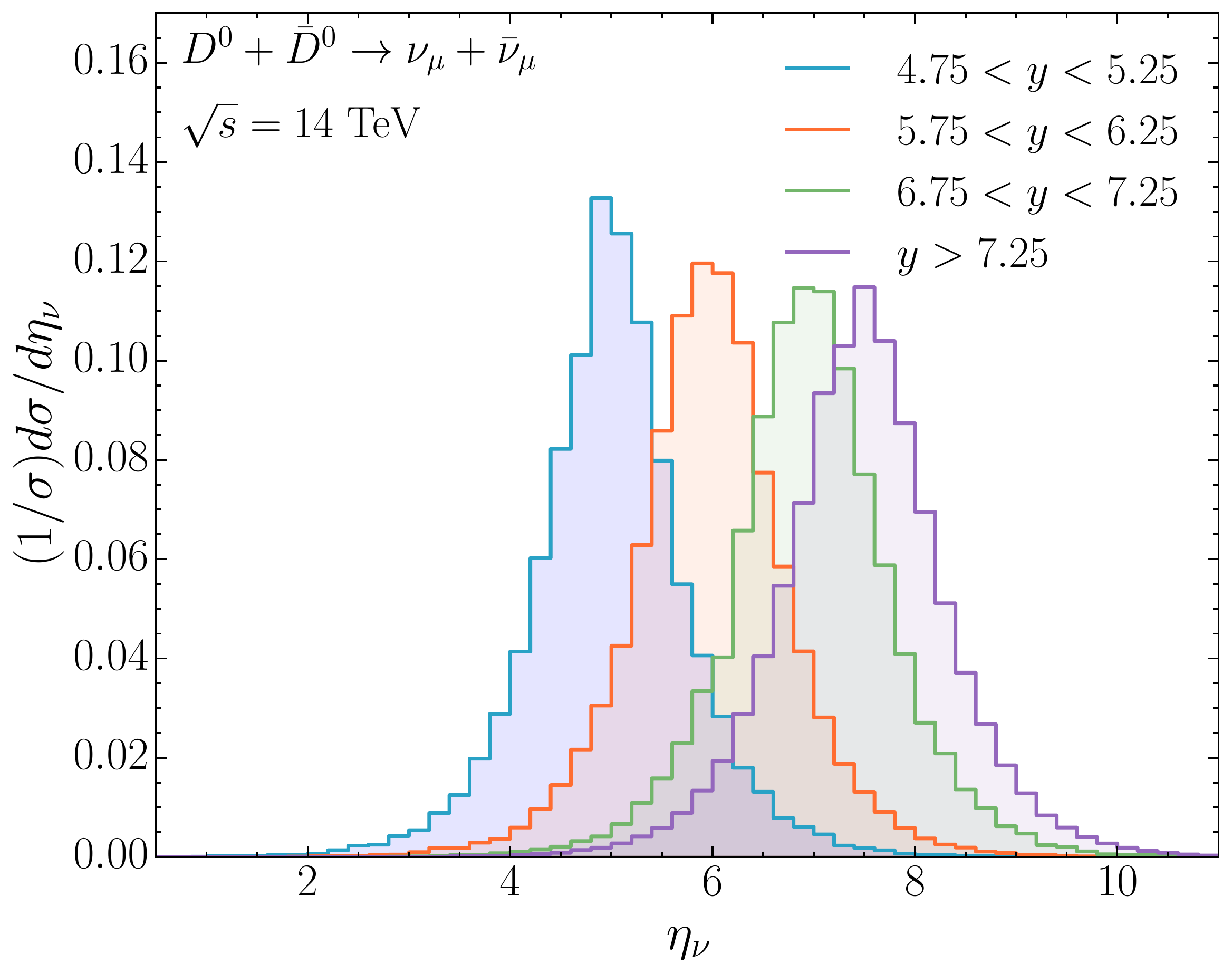}
          \includegraphics[width=.49\textwidth]{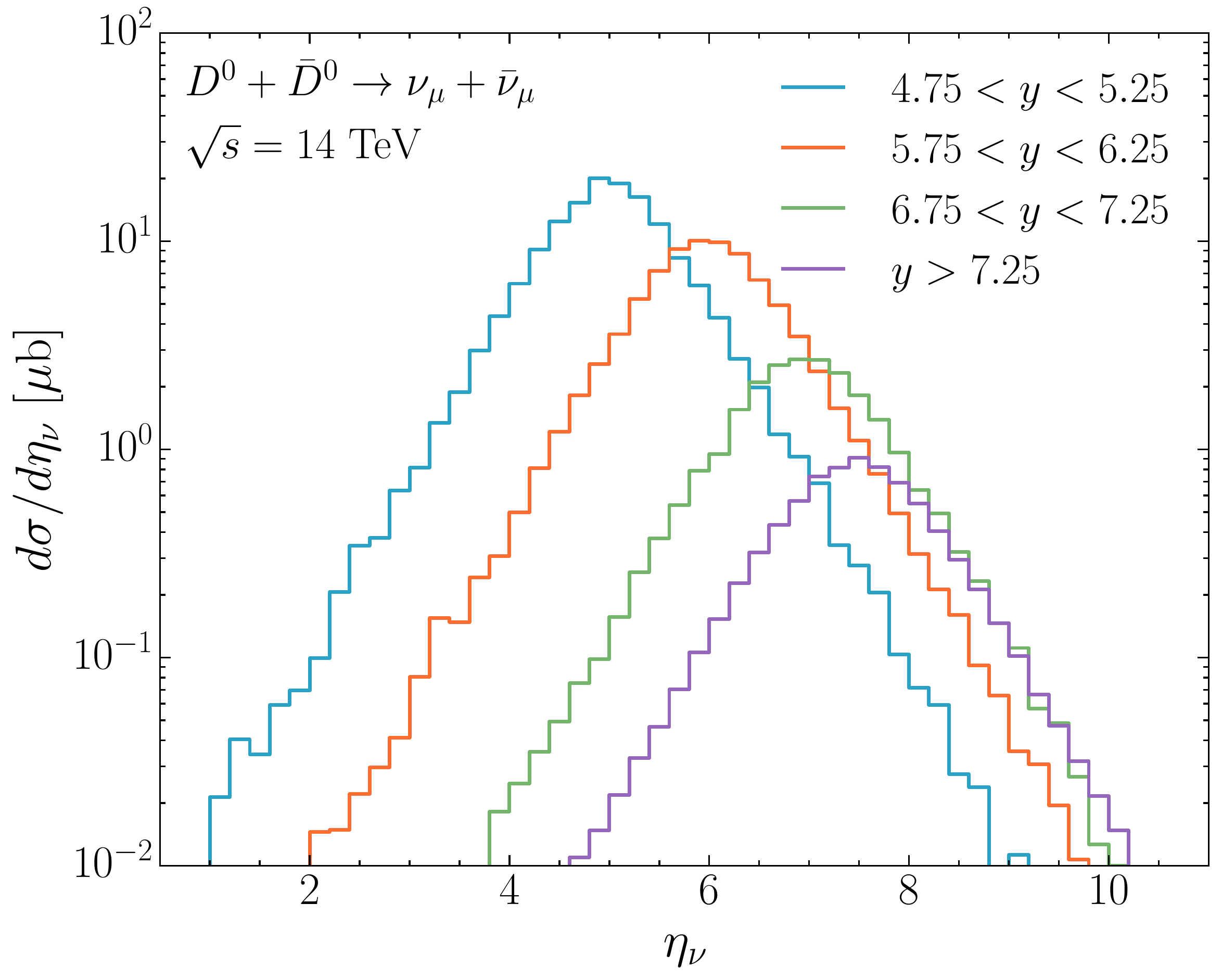}
             
       \caption{The cross-section normalized muon neutrino pseudorapidity distribution $(1/\sigma)d\sigma/d\eta_\nu$ (left) and the absolute muon neutrino pseudorapidity distribution $d\sigma/d\eta_\nu$ (right) for  $pp \to D^0+\bar{D}^0\to \nu_\mu+\bar\nu_\mu$ production from parent charm hadrons in rapidity ranges $4.75-5.25$, $5.75-6.25$, $6.75-7.25$ and for rapidity larger than $7.25$. Note the different scales on the vertical axis. 
       }
       \label{fig:dsdeta}
     \end{figure}

The forward neutrino experiments at the LHC investigate the region of $\eta_\nu > 7.2$. 
The neutrino rapidity is correlated with the charm hadron rapidity, as shown in the left panel of figure \ref{fig:dsdeta}. The left panel shows $(1/\sigma) d\sigma/d\eta_\nu$ for the $\nu_\mu+\bar\nu_\mu$ that come from decays of $D^0$ and $\bar{D}^0$ produced in hadron rapidity ranges $4.75 < y < 2.25$ (blue), $5.75 < y < 6.25$ (orange), $6.75 < y < 7.25$ (green) and $y>7.25$ (purple). The cross-section normalized distributions show spreads in $\eta_\nu$ centered around a narrow range of charm hadron $y$. The right panel of figure \ref{fig:dsdeta}, where absolute distributions are depicted, shows that in addition to the spread in $\eta_\nu$ relative to the parent-charm-hadron rapidities, the fact that
$d\sigma/d\eta_\nu$ falls quickly as $\eta_\nu$ increases means that for $\eta_\nu>8$ and 
$\sqrt{s}=14$ TeV, a nearly equal number of muon neutrinos from $D^0$ decays come from $D^0$ with $y>7.25$ and with $y$ in the range $6.75-7.25$. 

\begin{figure} 
 \centering
       \includegraphics[width=.49\textwidth]{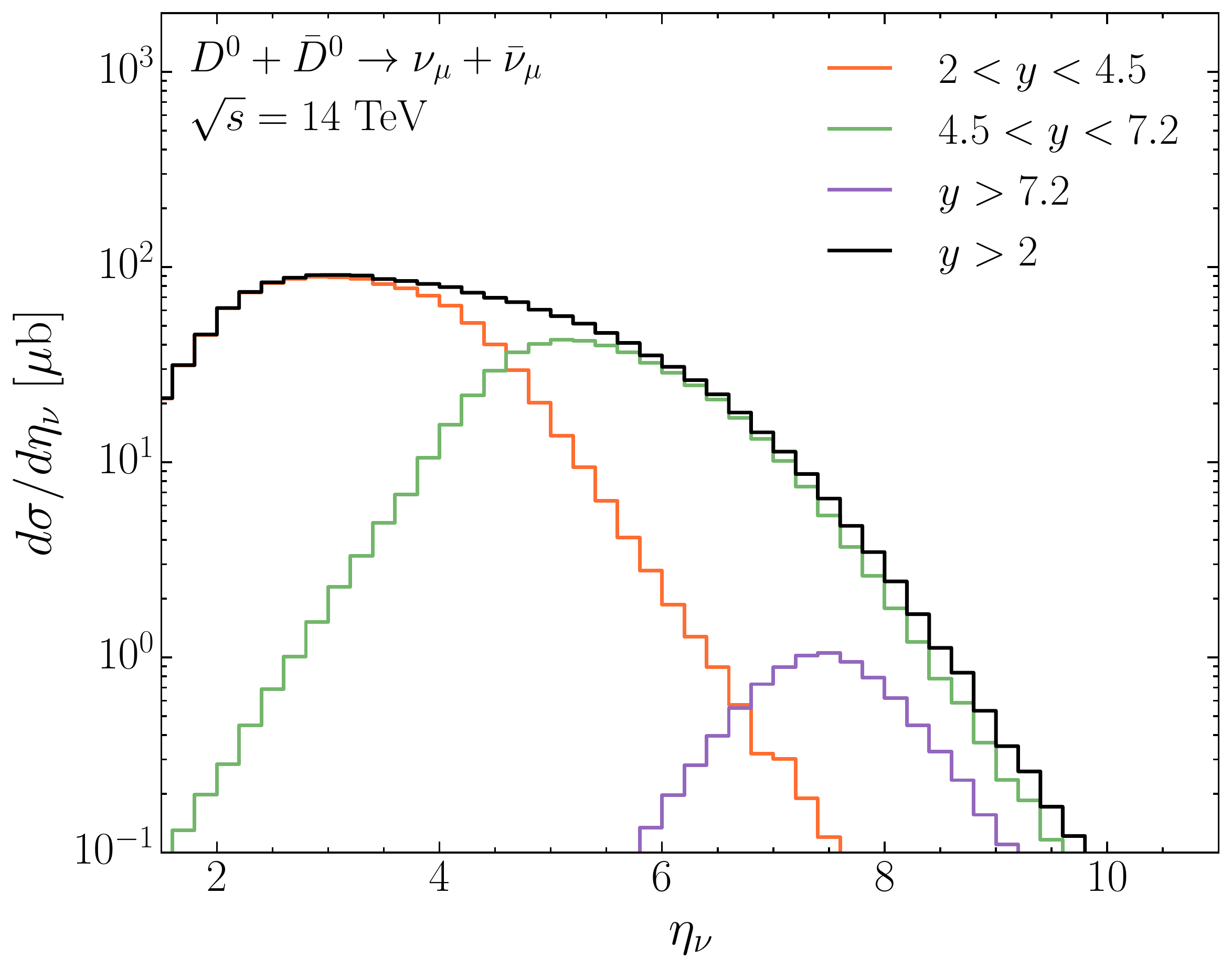}
       \includegraphics[width=.49\textwidth]{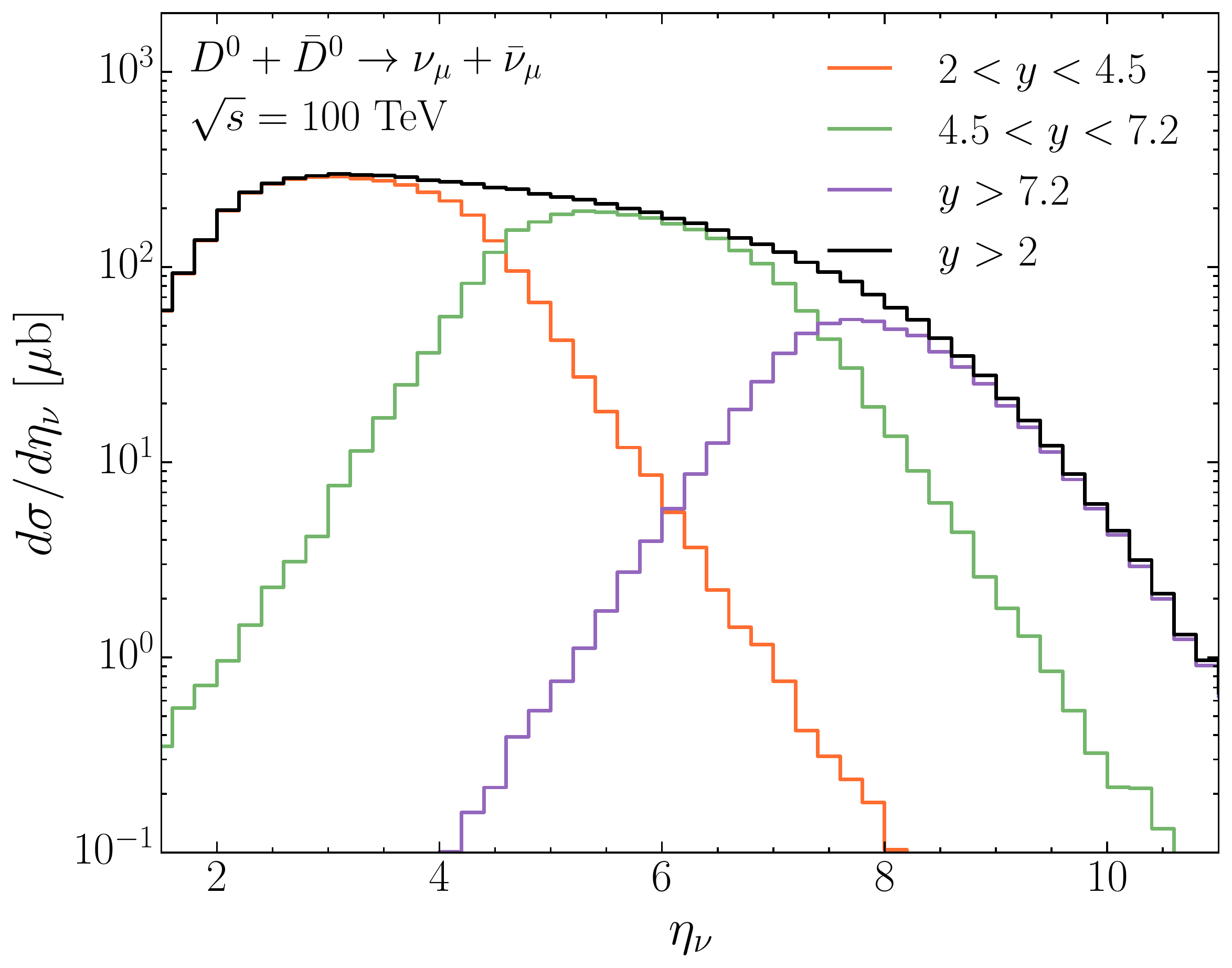}       
       \caption{
       The $\nu_\mu+\bar\nu_\mu$ pseudorapidity distribution from decays of $D^0 + \bar{D}^0$ produced in the rapidity ranges  $2<y<4.5$, $4.5<y<7.2$, $y>7.2$ and for $y>2$ in $pp$ collisions at $\sqrt{s}=14$ TeV (left) and $\sqrt{s}=100$ TeV (right). 
       } 
       \label{fig:dsdeta2}
     \end{figure}   
     
Figure \ref{fig:dsdeta2} shows the pseudorapidity distributions of muon neutrinos and antineutrinos from the $D^0$ and $\bar{D}^0$ produced in the rapidity ranges $2-4.5$, $4.5-7.2$ and greater than 7.2, as well as the sum ($y>2$), at a $pp$ collider with $\sqrt{s}=14$ TeV (left) and $\sqrt{s}=100$ TeV (right). The left panel of figure \ref{fig:dsdeta2} shows that 
the portion of $d\sigma/d\eta_\nu$ coming from the charm hadron rapidity range $4.5<y<7.2$ is actually larger than the one from $y>7.2$ at the LHC with $\sqrt{s}=14$~TeV. The ratio of the green histogram ($4.5<y<7.2$) to the purple histogram ($y>7.2$) is $\sim 2.9$ for $\eta_\nu=8$ and $\sim 2.1$ for $\eta_\nu=9$. This means that experiments at the Forward Physics Facility that measure neutrinos with $\eta_\nu>7.2$ will be sensitive to the rapidity range of charm-meson production that is important for the prompt atmospheric neutrino fluxes in the $\sim 10^6$~GeV energy range. 
     
The right panel of figure \ref{fig:dsdeta2} shows that for $\sqrt{s}=100$ TeV, the neutrino pseudorapidity distribution for $\eta_\nu\gtrsim 8$ is dominated by the decay of ($D^0$ + $\bar{D}^0$) produced with $y>7.2$, whose contribution is much larger than the one from
($D^0$ + $\bar{D}^0$) with $4.5<y<7.2$. For example, one can see that the purple histogram is a factor of $\sim 3.5(11)$ larger than the green histogram for $\eta_\nu=8(9)$.

\begin{figure}
 \centering
       \includegraphics[width=.49\textwidth]{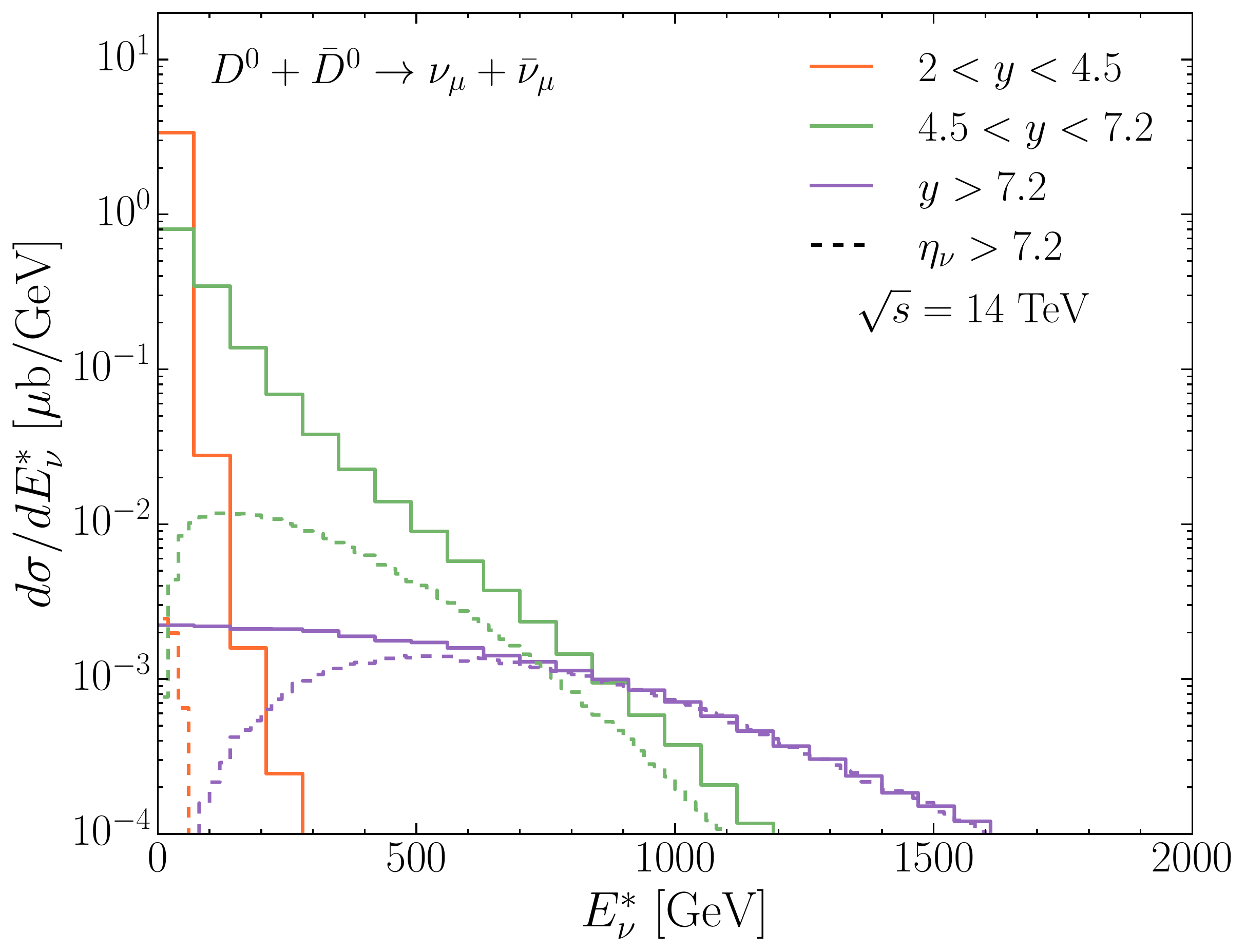}
       \includegraphics[width=.49\textwidth]{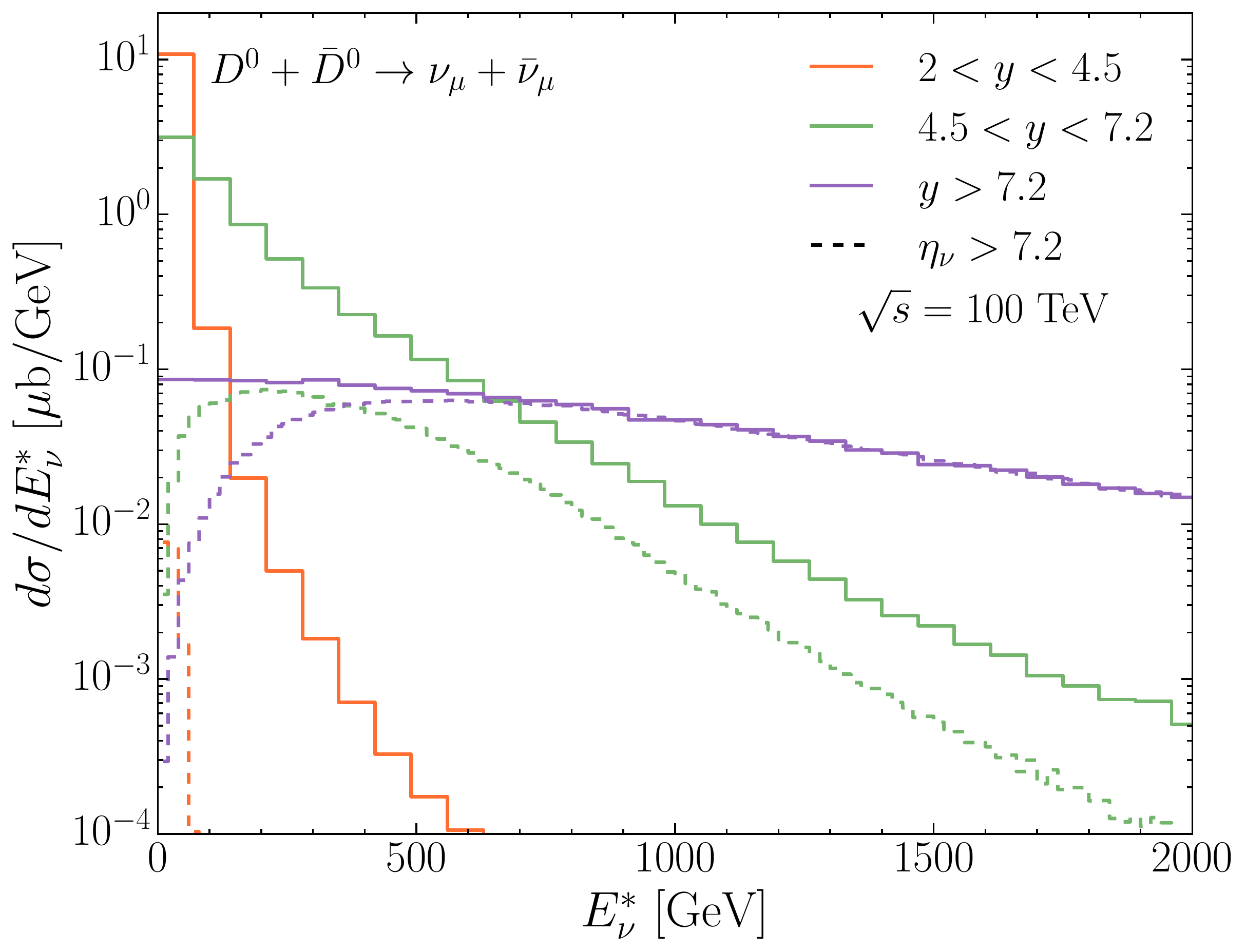}       
       \caption{The $\nu_\mu+\bar\nu_\mu$ collider energy $(E_\nu^*)$ distribution from $D^0$ and $\bar{D}^0$ in the rapidity ranges  $2<y<4.5$, $4.5<y<7.2$ and for $y>7.2$ in case of $pp$ collisions at $\sqrt{s}=14$ TeV (left) and $\sqrt{s}=100$ TeV (right). The solid histograms show $E_\nu^*$ distributions in the collider frame for all $\eta_\nu$, whereas all the dashed histograms show the energy distributions from the corresponding charm meson $y$ intervals that give neutrinos and antineutrinos with $\eta_\nu>7.2$, corresponding to the acceptance of detectors at the FPF. 
       } 
       \label{fig:dsdE2}
     \end{figure}     
   
To complete the discussion of the relation between the kinematic properties of $D^0$ and $\bar{D}^0$ at the LHC and the rapidities of neutrinos from charm in the prompt atmospheric neutrino fluxes, figure \ref{fig:dsdE2} shows the collider frame energy ($E_\nu^*$) distributions of neutrinos from $D^0+\bar{D}^0$ in the rapidity ranges  $2<y<4.5$, $4.5<y<7.2$ and for $y>7.2$ in case of $pp$ collisions at $\sqrt{s}=14$ TeV (left) and $\sqrt{s}=100$ TeV (right). The solid histograms show the neutrino energy distributions including all values of $\eta_\nu$. The dashed histograms show the energy distributions of neutrinos with $\eta_\nu>7.2$ from each of the aforementioned $D^0+\bar{D}^0$ rapidity ranges. 
The dashed histogram in the left panel shows that for charm hadron rapidity $4.5<y<7.2$, a substantial fraction of the neutrinos with energies above a few hundred GeV have $\eta_\nu>7.2$ (representative of an FPF detector coverage) for $\sqrt{s}=14$ TeV.
We note that the $\nu_\mu~+~\bar\nu_\mu$ energy distribution from $D^0 + \bar{D}^0$ essentially equals the $\nu_e+\bar\nu_e$ energy distribution from $D^0 + \bar{D}^0$. 
We also notice that the FPF detectors will be sensitive to a wide energy range, from $E^*_\nu$ of order tens of GeV to $E^*_\nu$ of the order of a few TeV.  
The $\nu_e+\bar\nu_e$ flux at the FPF is dominated by charm production and decay at high energy. In the 100's of GeV neutrino energy range, it will be important to disentangle the charm contribution from the kaon contribution to this flux.

At $\sqrt{s}=14$ TeV, corresponding to a proton beam energy in the fixed-target frame of $\sim 10^8$ GeV, the highest-energy collider neutrinos with $\eta_\nu>7.2$ come from charm hadrons produced with $y>7.2$, whereas at lower $E^*_\nu$ collider neutrinos come predominantly from hadrons produced with $y<7.2$ as shown in the left panel of figure \ref{fig:dsdE2}. 
The right panel of figure \ref{fig:dsdE2} shows that for $\sqrt{s}=100$ TeV, corresponding to a proton energy in the fixed-target frame of $\sim 5 \times 10^9$ GeV, there is a qualitatively similar behavior, but with an $E^*_\nu$ crossover point shifted to a lower value.

\subsection{Atmospheric neutrino fluxes from charm hadrons}

\begin{figure}
 \centering
       \includegraphics[width=.65\textwidth]{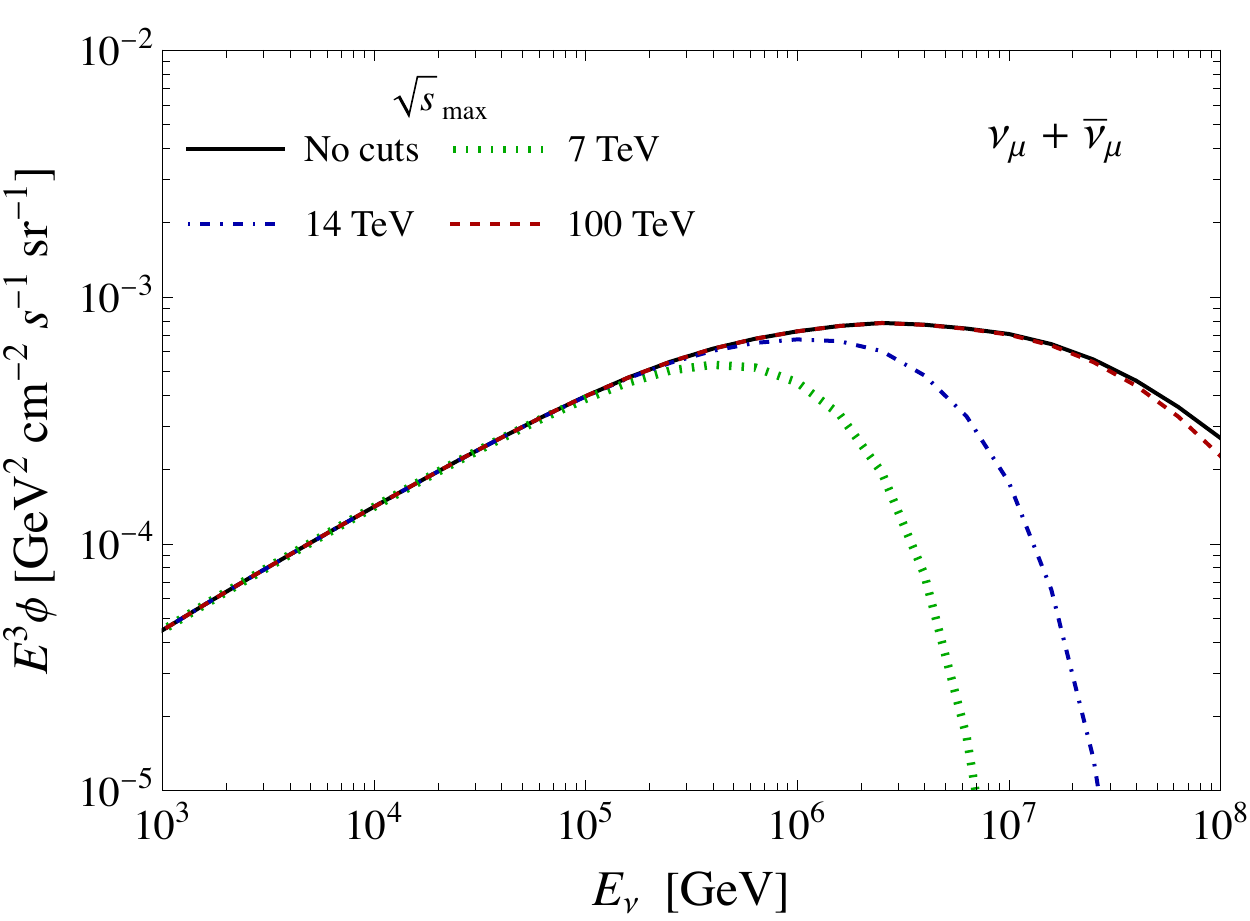}
   \caption{The prompt flux of atmospheric $\nu_\mu+\bar{\nu}_\mu$ for collision energies $\sqrt{s}$ varying up to different values of a maximum $\sqrt{s}_{\rm max}$.   }
       \label{fig:e3phi-rscut}
\end{figure}

\begin{figure}
    \centering
       \includegraphics[width=.49\textwidth]{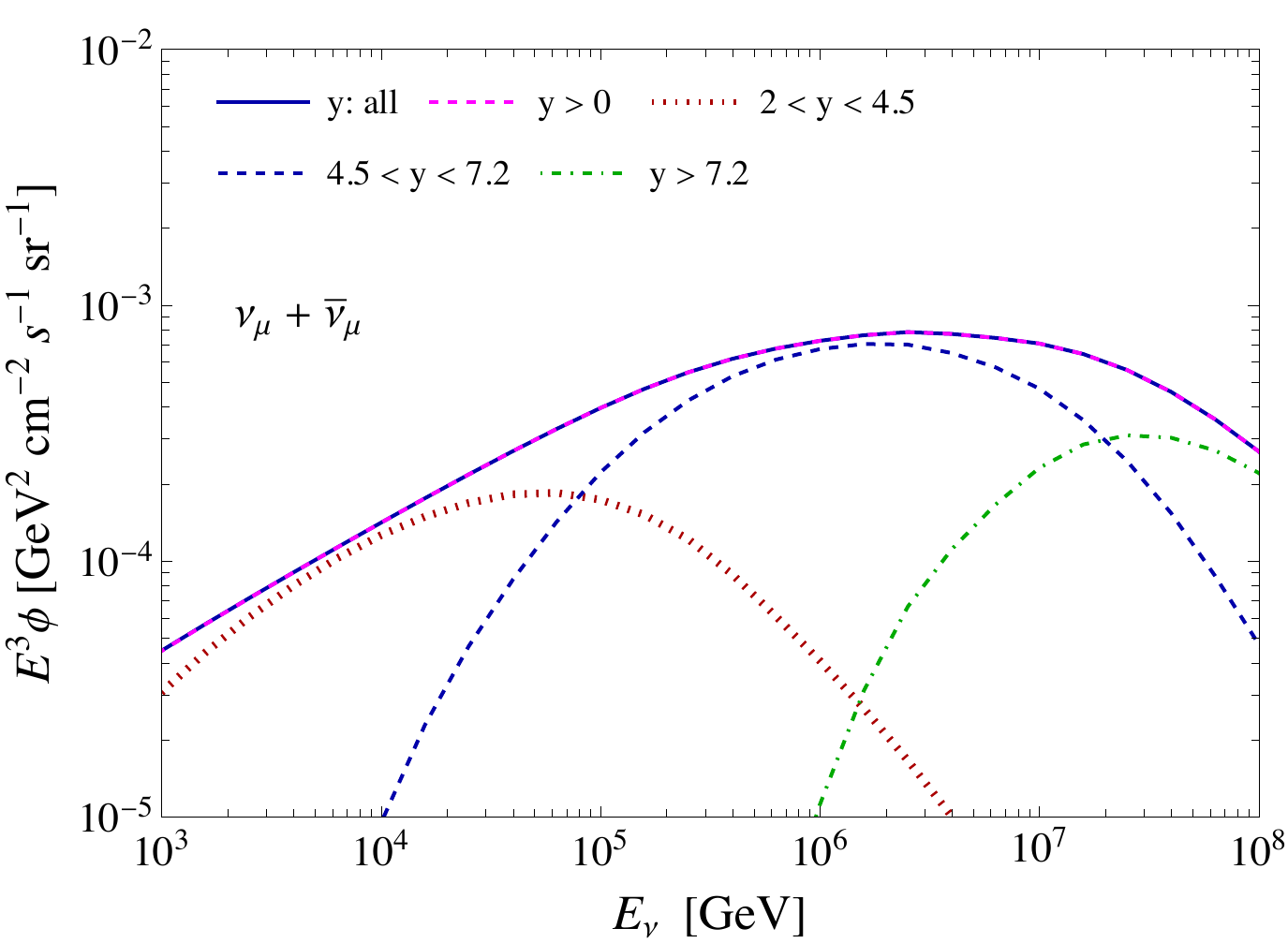}  
       \includegraphics[width=.49\textwidth]{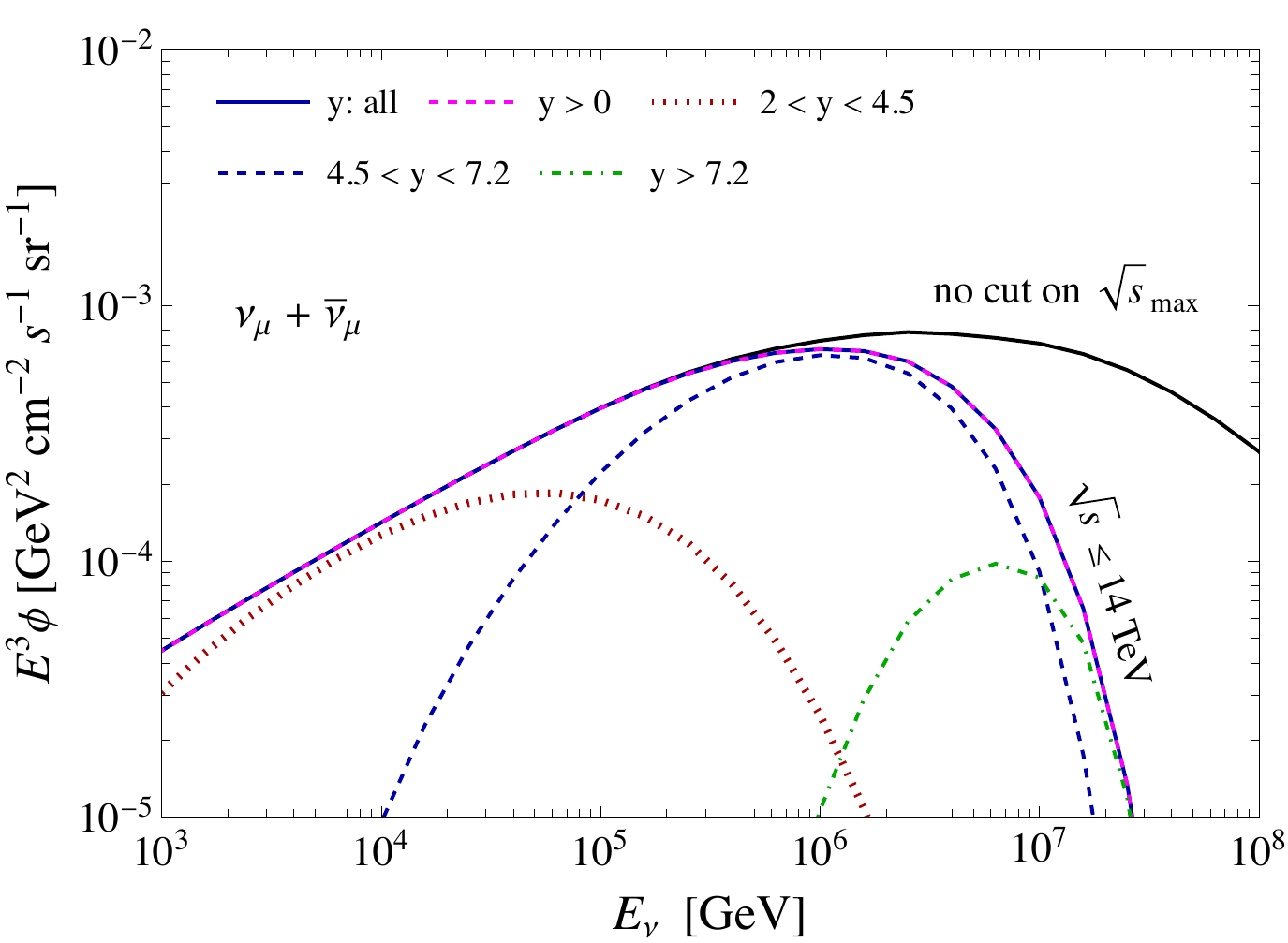}  
       \caption{The prompt flux of atmospheric $\nu_\mu+\bar{\nu}_\mu$ from different charm hadron  collider-frame rapidity ranges in
       $pp\to c\bar{c}X$, using a broken power law cosmic ray spectrum.
       The left panel includes the flux evaluated without cuts on $\sqrt{s}_{\rm max}$, whereas  the contributions in different rapidity ranges in the right panel are obtained under the condition $\sqrt{s}_{\rm max} = 14 {\ \rm TeV}$.}
       \label{fig:e3phi-ycut}
     \end{figure}

The resulting $\nu_\mu+\bar{\nu}_\mu$ fluxes of atmospheric neutrinos from the prompt decays of charm hadrons, scaled by $E_\nu^3$, are shown in Figs. \ref{fig:e3phi-rscut} and \ref{fig:e3phi-ycut}. We include contributions from the decays of  $D^0$, $D^+$, $D_s^+$ and $\Lambda_c^+$ and their antiparticles. 
For prompt neutrinos, the flux of $\nu_\mu+\bar{\nu}_\mu$ is essentially equal 
to that of $\nu_e+\bar{\nu}_e$ and is isotropic up to energies $E_\nu\sim 10^7$ GeV. Shown in Figs. \ref{fig:e3phi-rscut} and \ref{fig:e3phi-ycut} are the fluxes in vertical direction.

In figure \ref{fig:e3phi-rscut}, the predictions for different values of the maximum hadronic collision energy $\sqrt{s}_{\rm max}$ are presented, analogously to the predictions for the  $Z_{pD^0}$ in the left panel of figure \ref{fig:ZpD0_rs-Y}. 
As in case of the $Z_{pD^0}$-moment, appreciable contributions to the prompt neutrino flux evaluated with $\sqrt{s} \leq 14 {\, \rm TeV}$ appear below 10 PeV, in agreement with the findings of a previous study published in Ref.~\cite{Goncalves:2017lvq}.
The energies of produced neutrinos are lower than those of their parent particles, the fluxes of which are determined using $Z_{ph}(\sqrt{s}<\sqrt{s}_{\rm max})$.
The neutrino flux obtained under the cut  $\sqrt{s}_{\rm max} = 14~\tev$ accounts for 93\% and 25\% of the total flux evaluated with the whole range of CM hadronic-collision energies (no cuts) at $E_\nu = 10^6~\gev$ and $10^7~\gev$, respectively, while the same cuts on $\sqrt{s}_{\rm max}$ yield 99\% and 63\% of contributions to $Z_{pD^0}$ at the corresponding energies.
Considering that the prompt atmospheric neutrino flux is dominant over the conventional atmospheric neutrino flux above $E_\nu = 10^5 - 10^6 \ \gev$, we can conclude that the study of charm-hadron production at the LHC will be able to constrain the prompt atmospheric neutrino flux in the transition region and up to $E_\nu \lesssim 10^7 \ \gev$. 
On the other hand, as expected from 
figure~\ref{fig:e3phi-rscut}, the charm hadrons produced at $\sqrt{s} \leq 100 {\, \rm TeV}$ impact the full neutrino energy range relevant for IceCube \cite{IceCube:2020wum} and IceCube-Gen2 \cite{IceCube-Gen2:2020qha} sensitivity to the prompt atmospheric neutrino component. This means that the FCC will be able to provide precious data to further improve the theoretical predictions of the high-energy prompt-neutrino fluxes.

In figure \ref{fig:e3phi-ycut}, we show the atmospheric fluxes of $\nu_\mu+\bar{\nu}_\mu$ arising from  charm mesons produced in different center-of-mass rapidity ranges. 
The predictions in the left panel are evaluated with full range of $\sqrt{s}$, while the ones in the right panel are obtained under the cut $\sqrt{s} \leq 14 {\ \rm TeV}$.
The predictions in figure \ref{fig:e3phi-ycut} indicate that the prompt atmospheric neutrinos with energies $E_\nu \gtrsim 10^5~\gev$, where their flux is important relative to the conventional atmospheric and the diffuse astrophysical neutrino fluxes, come mainly from charm hadrons produced with  CM rapidities in the range of $y \gtrsim 4.5$. 

The LHC Run 3 forward neutrino experiments SND@LHC and FASER$\nu$ cover the pseudo-rapidities $7.2 < \eta_\nu < 8.6$ and $\eta_\nu \gtrsim 8.5$, respectively \cite{Feng:2022inv, VanHerwijnen:2022sih}.
The Forward Physics Facility at the HL-LHC is foreseen to host three neutrino detectors, i.e. the Advanced SND (AdvSND) FAR detector, FASER$\nu$2 and FLArE, with the respective coverage of $7.2 < \eta_\nu < 9.2$, $\eta_\nu \gtrsim 8.5$ and $\eta_\nu \gtrsim 7.5$ \cite{Feng:2022inv}. These detectors are still under design, so there is the possibility to cover other pseudo-rapidity ranges.
On the other hand, the AdvSND NEAR detector, located in another area of LHC, will cover $4<\eta_\nu<5$. 
We have shown  in figure \ref{fig:dsdeta2} that for $\sqrt{s}=14$ TeV, charm hadrons produced in the range $4.5<y<7.2$ can yield neutrinos with $\eta_\nu>7.2$. Thus, all of the forward neutrino experiments at the LHC are sensitive to the kinematic regions relevant for the prompt atmospheric neutrino flux for $E_\nu\gtrsim 10^5$ GeV.

For reference, we also evaluated the atmospheric neutrino flux from charm hadrons that come from CM rapidity $y>8.5$, namely, neutrinos from charm hadrons that, if produced in the collider, would have momenta that point in the direction of the  FASER$\nu$/FASER$\nu$2 detectors. 
For the atmospheric neutrino fluxes evaluated under the cut $\sqrt{s}_{\rm max} = 14~\tev$, the contribution from charm hadron CM rapidities $y > 8.5$ to the total flux is much less than 0.1\% at $E_\nu = 10^7~\gev$, therefore completely negligible. Even in case calculated with no cut on $\sqrt{s}$, the contribution from charm hadrons with CM rapidity $y > 8.5$ is less than 1\% at $E_\nu = 10^7~\gev$, and at most $\sim$ 20\% at $E_\nu = 10^8~\gev$.

 \begin{figure} 
    \centering
       \includegraphics[width=.45\textwidth]{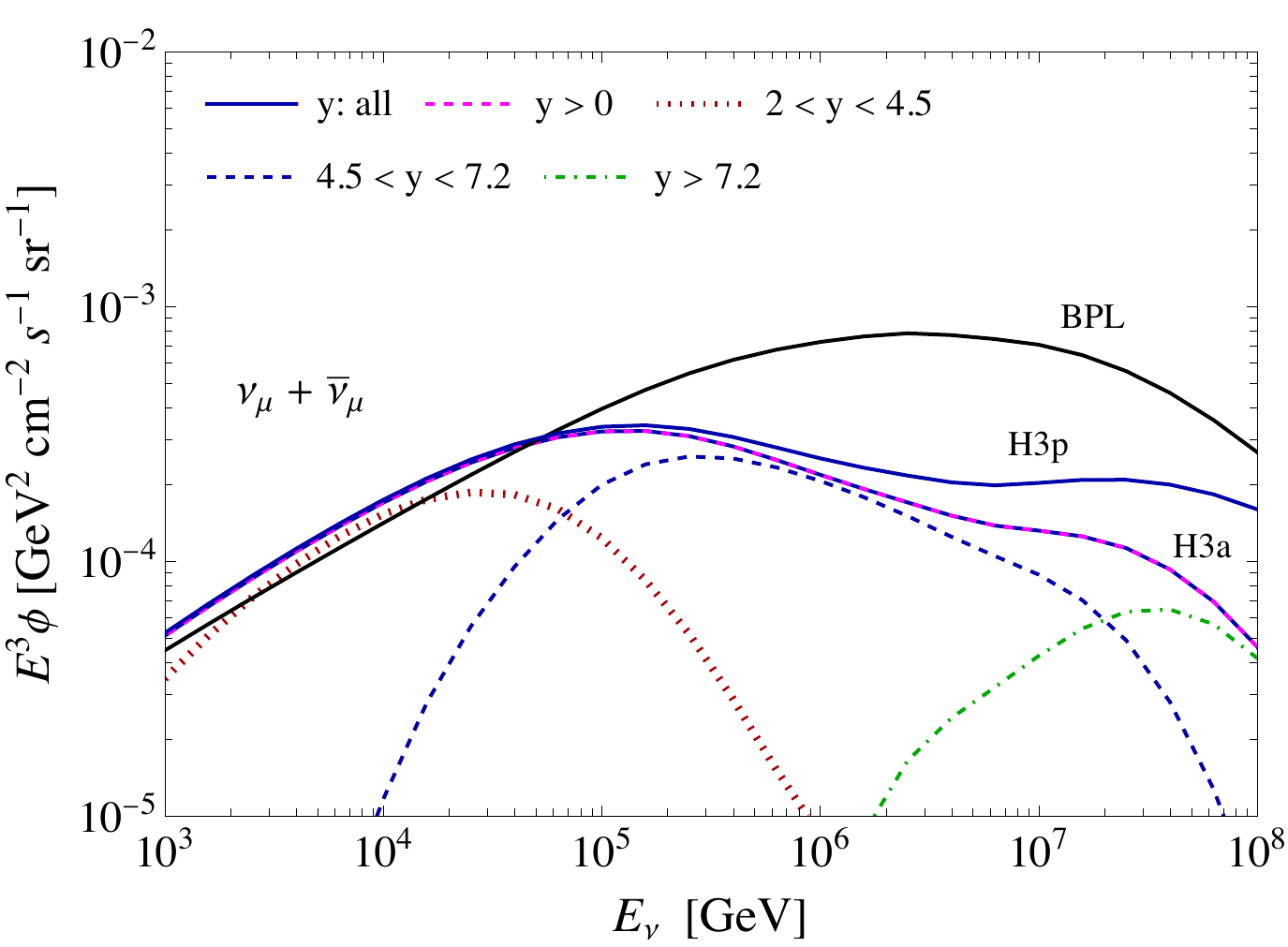} 
       \includegraphics[width=.45\textwidth]{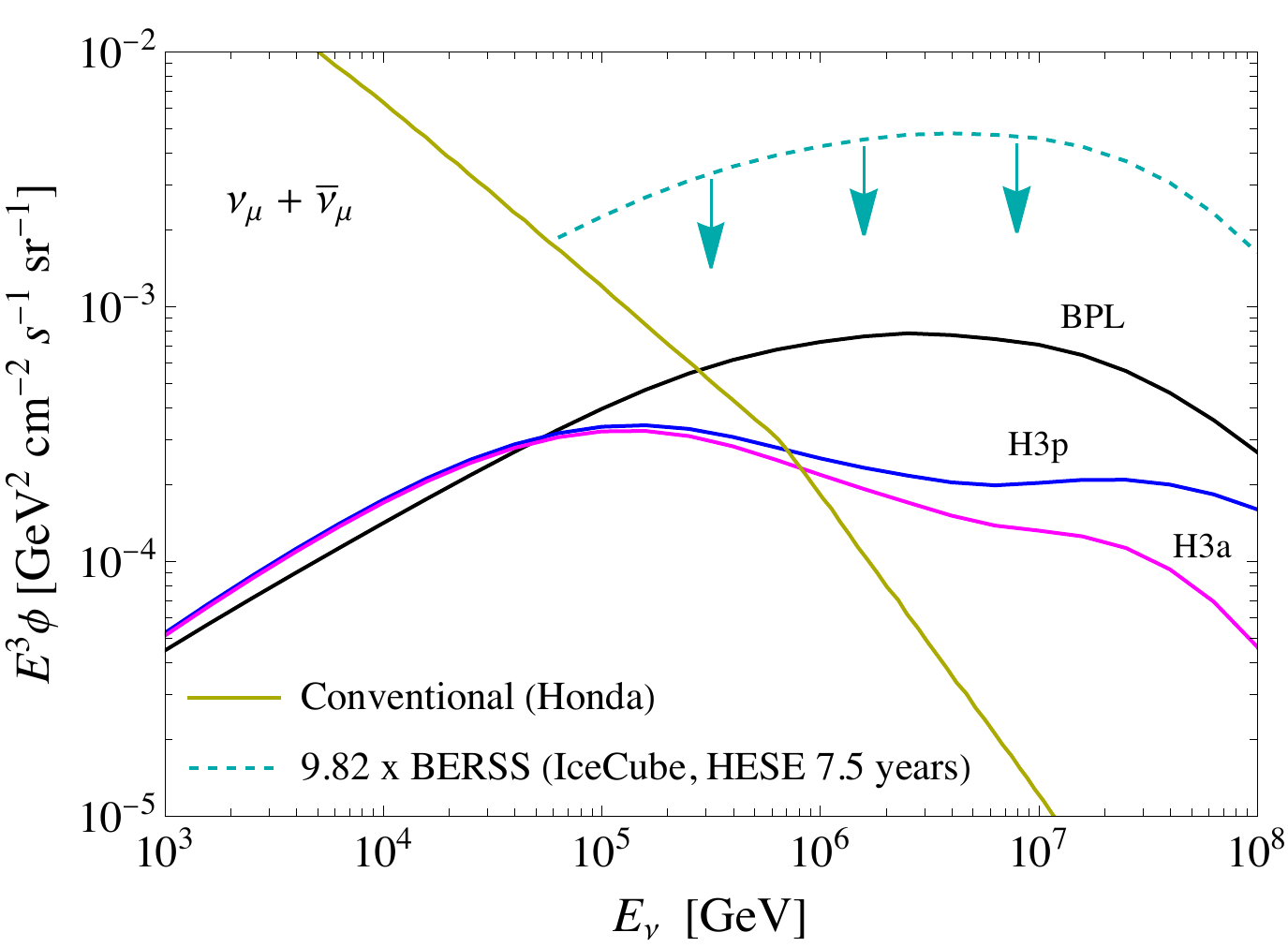}       
       \caption{Left: The prompt flux of atmospheric $\nu_\mu+\bar{\nu}_\mu$ from different charm meson CM rapidity ranges in
       $pp\to c\bar{c}X$ using the H3p and H3a spectra as well as the broken power law  spectrum for cosmic ray fluxes. Right: 
       The prompt flux of atmospheric $\nu_\mu+\bar{\nu}_\mu$ from H3p, H3a and BPL cosmic ray spectra is shown with the vertical conventional $\nu_\mu+\bar{\nu}_\mu$ flux
       \cite{Honda:2006qj} and the IceCube upper limit on the prompt atmospheric flux \cite{IceCube:2020wum}, expressed as a scaling of the BPL flux from ref. \cite{Bhattacharya:2015jpa}. 
}
       \label{fig:e3phi-CR}
     \end{figure}

Thus far, we have used the BPL cosmic-ray spectrum to illustrate 
the impact of hadronic collisions at different $\sqrt{s}$ and charmed mesons produced in different CM rapidities  
on the prompt atmospheric neutrino flux. 
For reference, the left panel of 
figure \ref{fig:e3phi-CR} presents the prompt fluxes of atmospheric $\nu_\mu+\bar{\nu}_\mu$ evaluated with the H3p and H3a cosmic ray spectra in addition to the one with the BPL spectrum. 
 Similarly to the left panel of figure~\ref{fig:e3phi-ycut}, the contributions from the different charm hadron CM rapidity ranges 
 are presented for the predictions evaluated with the H3a cosmic-ray spectrum. The conclusions are similar to those already drawn when analyzing the BPL case. 

Finally, the right panel of figure \ref{fig:e3phi-CR} shows the 
vertical prompt atmospheric $\nu_\mu+\bar{\nu}_\mu$ fluxes along with the vertical conventional atmospheric neutrino flux \cite{Honda:2006qj} and the IceCube Collaboration's 90\% upper limit on the prompt $\nu_\mu+\bar{\nu}_\mu$ flux from the analysis of high energy starting events (HESE) collected in 7.5 years \cite{IceCube:2020wum}. The upper limit is represented by a scaling of the BERSS prediction \cite{Bhattacharya:2015jpa}. 
A comparison between the left and right panels of figure \ref{fig:e3phi-CR} shows that in the energy range where the transition from the conventional atmospheric neutrino flux to the prompt atmospheric neutrino flux occurs,  prompt neutrinos from charm hadron production with equivalent CM rapidities $4.5<y<7.2$ are most important.
The reader can refer to ref.~\cite{Jeong:2023gla} (and section 8.2 of ref.~\cite{Feng:2022inv}) for a comparison of these results with those from other groups \cite{Garzelli:2015psa,Bhattacharya:2016jce,Gauld:2015kvh, 
Bhattacharya:2015jpa,Enberg:2008te,Goncalves:2017lvq,Fedynitch:2018cbl,Zenaiev:2019ktw}.
Compared to the BERSS prediction, our evaluation is larger by a factor of 1.5 -- 1.8. Modifications in the PDF, QCD scale choice and the implementation of fragmentation for charm hadron production are cause of the differences. QCD scale uncertainties are large and encompass a wide range of predictions.
We expect that our conclusions concerning kinematic regions of charm production relevant for prompt neutrino production also apply to the predictions of the other prompt neutrino groups.

\section{Uncertainty from the parton distribution functions}

\begin{figure} 
     \centering
       \includegraphics[width=.49\textwidth]{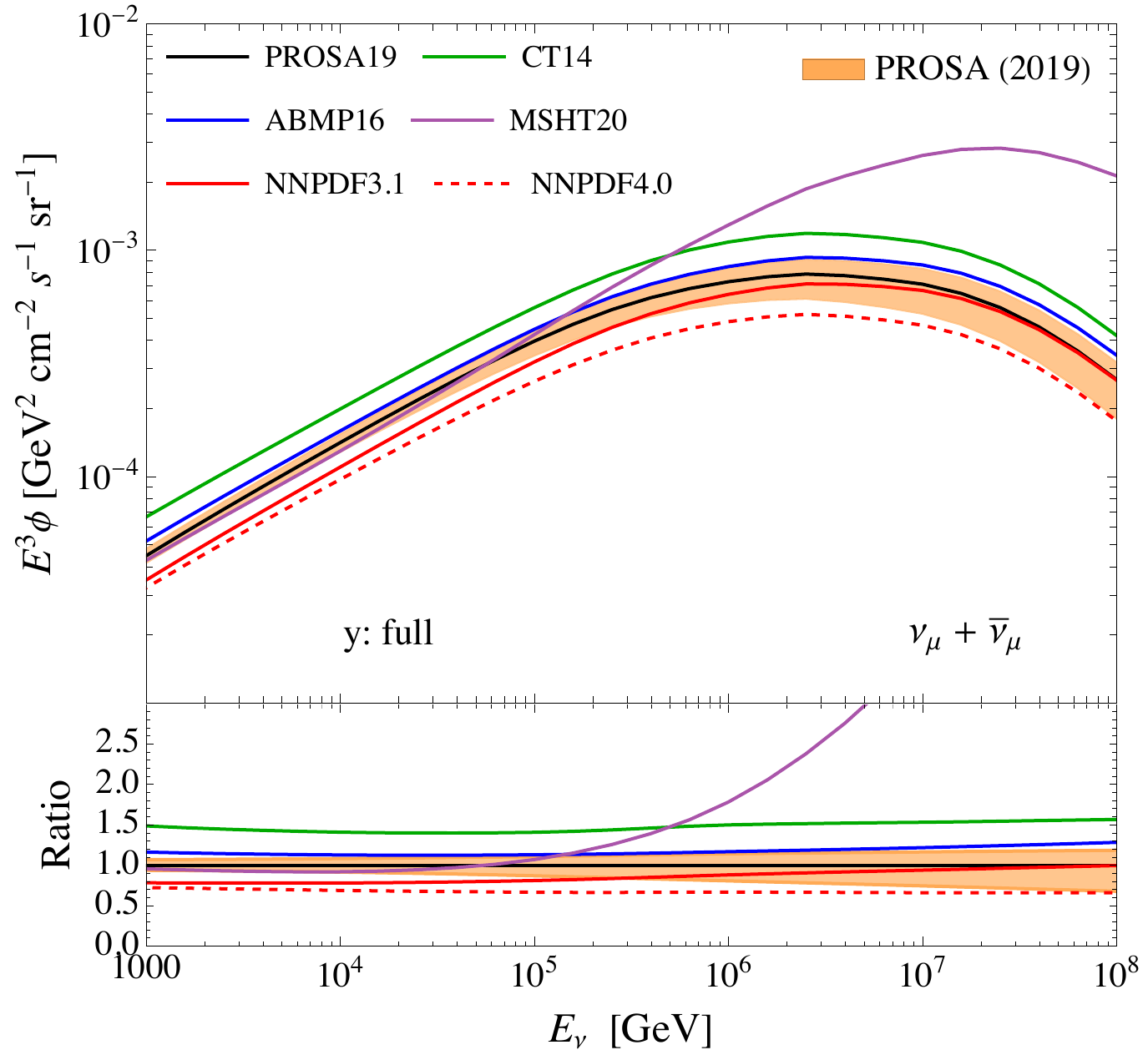}
       \includegraphics[width=.49\textwidth]{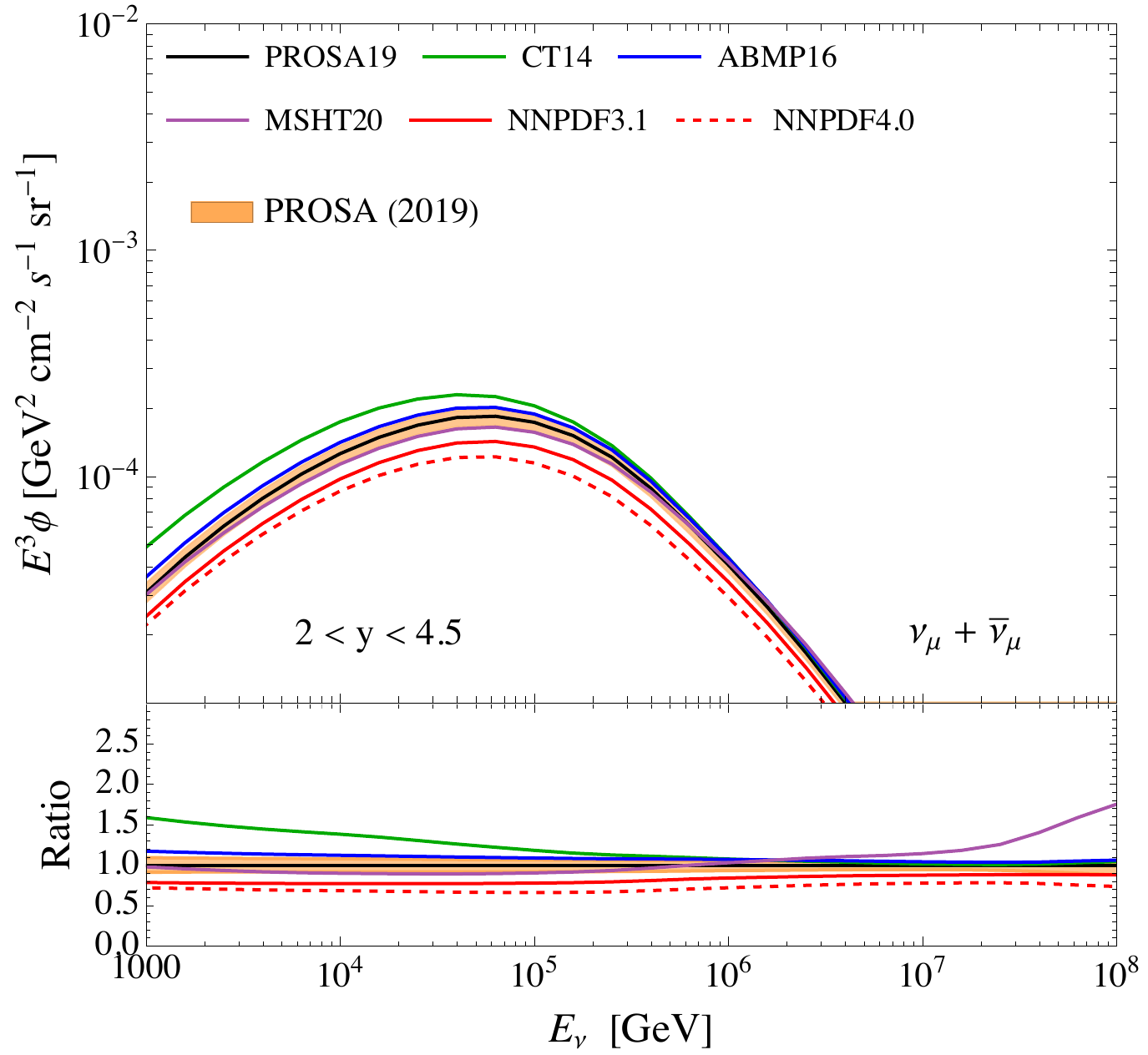}
       \includegraphics[width=.49\textwidth]{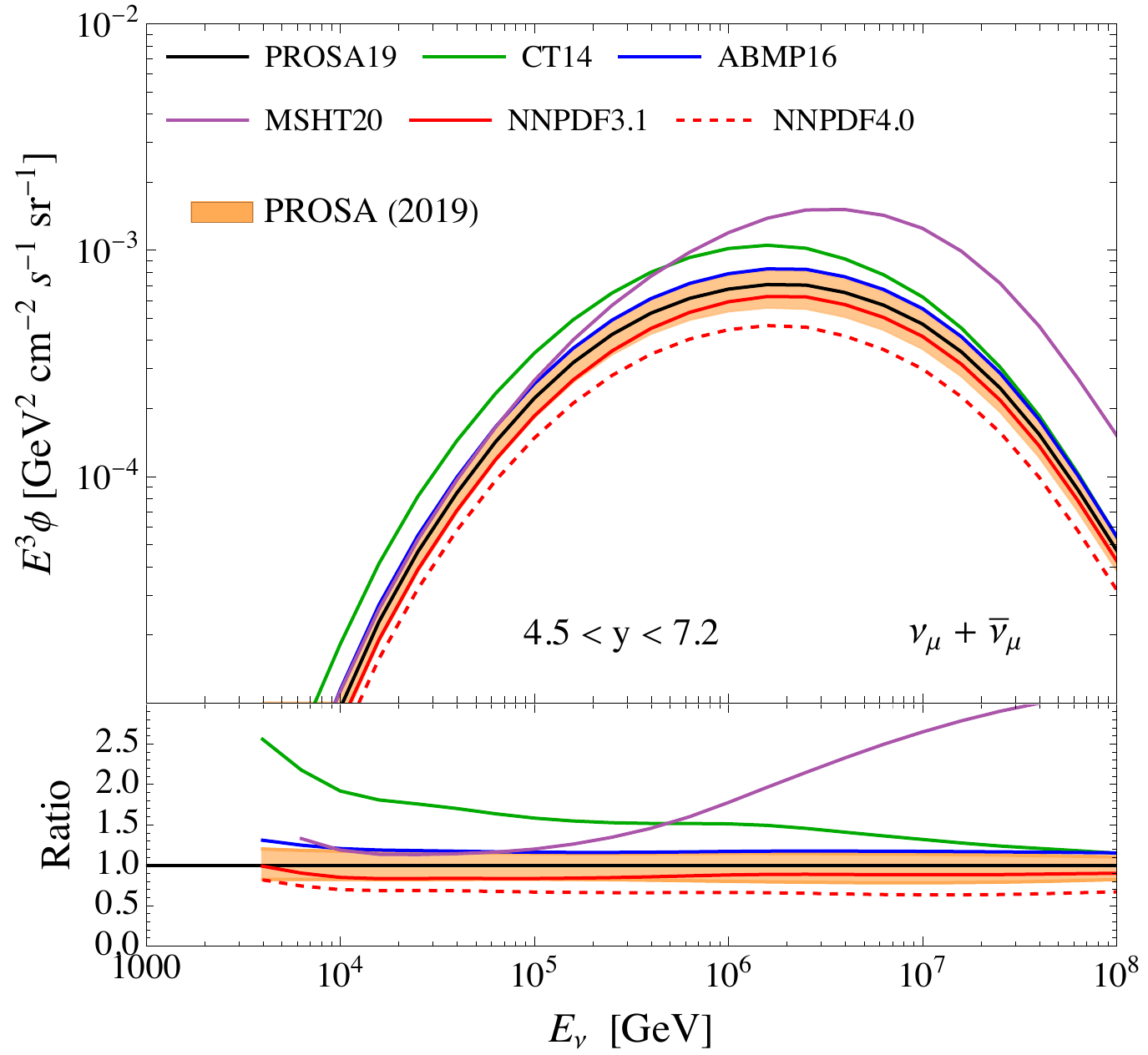}
       \includegraphics[width=.49\textwidth]{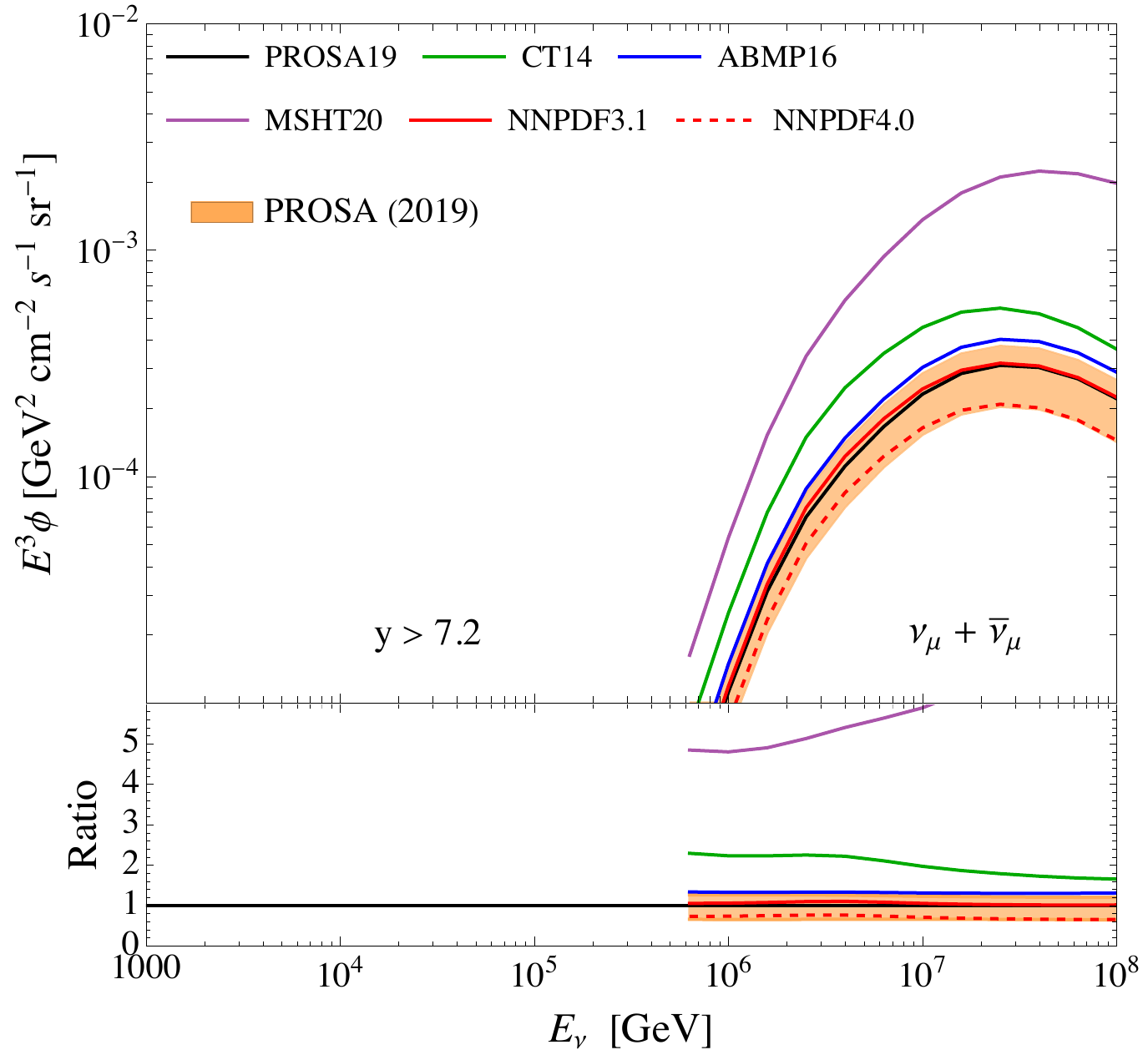}  
       \caption{
       The prompt atmospheric $\nu_\mu+\bar{\nu}_\mu$ fluxes with uncertainties from the 40 eigenvectors of the PROSA 2019 PDF fit  and with the central PDFs by other groups presented in figure \ref{fig:pdf-range}. 
       The contributions of neutrinos from charm hadrons in different CM rapidity ranges are shown in different panels.
       The ratios of predictions to the central PROSA PDFs are presented in the lower part of each panel. 
       }
       \label{fig:e3phi-x}
     \end{figure}

As discussed in the previous section, for neutrinos produced in the forward region, the colliding partons in the interaction have asymmetric longitudinal momentum fractions ($x$). High-energy neutrinos from charm produced in interactions of cosmic rays with a steeply-falling energy spectrum favor a momentum fraction from cosmic rays that is large, while the momentum fraction of the parton in the struck air nucleon is very small. At the CM energies relevant for this work, the respective values of $x$ can be larger than 0.1 and less than $10^{-5}$. As already discussed in Section 3.1, the PDFs in such regions have been poorly constrained due to lack of experimental data, with the exception of LHCb data for heavy-flavor production.
For $E_\nu \gtrsim 10^6~\gev$, the contribution of collisions with gluons in the nucleon target with $x < 10^{-5}$ is significant. 
At higher energies, the lower $x$ region become even more important, which is reflected by the trends in the atmospheric neutrino flux contributions from different rapidity regions in figure~\ref{fig:e3phi-x}.
On the other hand, the effect of variations in the large-$x$ PDFs 
affects the normalization of the prompt neutrino flux for the whole energy range, and contributions from PDFs in the range of $x>0.6$ have a negligible effect~\cite{Goncalves:2017lvq}. 

Using the BPL cosmic-ray flux, we illustrate the energy and CM-rapidity dependence of the PDF uncertainties associated with the PROSA PDF fit procedure, theoretical model and parameterizations on the predicted prompt atmospheric neutrino flux in figure \ref{fig:e3phi-x}. The orange band is the uncertainty envelope determined using the 40 variants in the PROSA PDF fit (see appendix~A of ref. \cite{Bai:2021ira}). Each of the four panels of figure~\ref{fig:e3phi-x} also shows results obtained with central NLO PDFs from the CT14, ABMP16, MSHT20 and NNPDF3.1 fits as well as the newer NNPDF4.0 version. Each of the PDFs have associated uncertainties, but we only show the PROSA PDF uncertainties in this work.
The upper left panel shows the prompt flux and uncertainty band for 
the full CM-rapidity range relevant for prompt neutrino production, along with the ratio of the predictions with different PDFs to those with the central PROSA PDFs. The three other panels of figure \ref{fig:e3phi-x} show the results for the restricted CM-rapidity ranges of $2<y<4.5$, $4.5<y<7.2$ and $y>7.2$. 

The results in figure \ref{fig:e3phi-x} show that the overall uncertainty of the PROSA PDF predictions increases with energy, and it is within $+25$\% and $-35$\% around the central prediction for $E_\nu \lesssim 10^8~\gev$. The fluxes for $2 < y < 4.5$  dominantly contribute to the total flux below $\sim 10^5~\gev$, where the uncertainty due to the different PROSA PDF eigenvectors are within $\pm 10\%$.
The LHCb data on heavy-flavour production in this rapidity range are included in the PROSA PDF fit \cite{Zenaiev:2019ktw}. 
The different eigenvectors of the PROSA PDF fit yield an uncertainty in the range of $+15\%$ and $-20\%$ for the fluxes with $4.5 < y < 7.2$, which are dominant for the energy range of $10^5~\gev \lesssim E_\nu \lesssim 10^7~\gev$.  The PROSA PDF uncertainty increase to $+25\%$ and $-35\%$ at $10^7~\gev$ for the fluxes from charm hadrons with CM rapidities $y > 7.2$. 

In the region where experimental data are not available, the uncertainty from the PROSA PDFs 
strongly depend on the parameterization chosen,
and correspond to an extrapolation, although 
still constrained by sum rules.  
Comparing with the results evaluated using the other PDFs considered here, one can see that the fluxes with the CT14 and MSHT20 central PDF sets are out of the PROSA uncertainty band (but they would still at least partially overlap if one would draw the CT14 and MSHT20 uncertainties). The CT14 central PDFs yield the large fluxes with difference of $40 - 60\%$ with respect to the central prediction with the PROSA PDFs for $10^3~\gev \leq E_\nu \leq 10^8~\gev$. This is related to the large-$x$ behaviour of the considered PDFs.  On the other hand, the MSHT20 fluxes differ by as much as a factor of 3 for the contribution from $4.5<y<7.2$ and by as much as a factor of 9 for $y>7.2$, as shown in the lower panels of figure \ref{fig:e3phi-x}. This is attributed to the dramatic increase of the gluon distribution as $x$ decreases below $x\sim 10^{-5}$ (see figure~1).
The ABMP16 central predictions lie almost on the upper boundary of the PROSA uncertainty band in all  rapidity ranges. 
The evaluations with the NNPDF central NLO PDF sets are mostly lower than the PROSA predictions.
For $E_\nu < 10^5~\gev$ and $2 < y < 4.5$, both NNPDF3.1 and NNPDF4.0 central predictions are even smaller than the lower boundary of the PROSA range.
While the NNPDF4.0 PDFs yield fluxes that differ from the central predictions of PROSA by about 30\% all over the presented energies, in case of the NNPDF 3.1 PDFs, the discrepancy is reduced as the energy increases, and both predictions eventually converge towards each other at very high energies, i.e., near $E_\nu \sim 10^8~\gev$. 
This can also be seen in the panels for the different rapidity ranges. 
For $2<y<4.5$, where the contribution to the total fluxes dominate at low energies $E < 10^5~\gev$, both NNPDF central predictions are placed out of the PROSA uncertainty band, but would overlap with it if one would consider the NNPDF uncertainty bands.  
On the other hand, for $y>7.2$, which impact the prompt flux for $E_\nu \sim 10^7~\gev$, the predictions with central NNPDF3.1 and PROSA PDFs turn out to be consistent with each other  within 5\%. 

As outlined at the beginning of Sec. 4, the prompt atmospheric neutrino fluxes are only moderately sensitive to the behaviour of PDFs at very large $x$ ($x\gtrsim 0.6$). On the other hand, as shown in ref. \cite{Bai:2021ira}, the prompt neutrino flux at the TeV energy scale at the FPF is very sensitive to PDFs at very large $x$. FPF neutrino energy distributions for collider $E_\nu<1$~TeV are less sensitive to the very large-$x$ PDFs. It is the latter energy region for collider neutrinos (as shown in figure \ref{fig:dsdE2}) that is most relevant to the prompt atmospheric flux energy range of interest. While there are differences in normalizations and shapes of the contributions from the three rapidity ranges to the prompt atmospheric flux in figure \ref{fig:e3phi-x}, all of the PDF evaluations show the same qualitative behavior, namely that the $4.5<y<7.2$ range for charm hadron rapidity is the most important for the prompt atmospheric neutrino flux in the laboratory energy range of $E_\nu=10^5-10^7$ GeV.

\section{Discussion and conclusions}

In this work, we have investigated 
relevant kinematic regions for atmospheric prompt neutrino production in terms of $\sqrt{s}$ of the colliding nucleon-nucleon system and the CM rapidity $y$ of the charm hadrons decaying into neutrinos. 
Our focus includes the fixed-target energy range from $E_\nu\sim 10^4$~GeV where the conventional atmospheric neutrino flux is larger 
than the prompt atmospheric neutrino flux,  to $E_\nu\sim 10^8$ GeV, where the prompt neutrino flux dominates over the conventional neutrino flux and the astrophysical neutrino flux is relevant \cite{IceCube:2013cdw,IceCube:2020acn,IceCube:2020wum}.
As shown in Figs.~\ref{fig:e3phi-rscut} and \ref{fig:e3phi-ycut}, the LHC $\sqrt{s}$ does not fully cover this neutrino energy range.
In fact, the maximum CM energy of
$pp$ collisions at the LHC $\sqrt{s} = 14~\tev$ corresponds to $E_p\sim 10^8~\gev$ in a fixed-target frame.
A fraction of the cosmic-ray nucleon energy is transferred to charmed mesons, and neutrinos also take a fraction of the energy of the decaying charmed mesons. As the figures show, prompt atmospheric neutrinos from nucleon-nucleon collisions with center-of-mass energy
$\sqrt{s} \leq 14$ TeV are distributed in a fixed-target energy range of $E_\nu\lesssim 10^7~\gev$.  FCC running in hadron-hadron collision mode with $\sqrt{s}=100$ TeV would directly connect to prompt atmospheric neutrinos with energies $E_\nu = 10^7~\gev$ and higher.

Although the direct correspondence between LHC production of neutrinos from charm and cosmic ray production of neutrinos from charm is limited to atmospheric neutrino energies $E_\nu \lesssim 10^7~\gev$, this covers the energy ranges where the transition   from the conventional atmospheric neutrino flux
to the prompt one occurs and the astrophysical flux becomes dominant. LHC forward experiments can play important roles in improving theoretical predictions of prompt atmospheric neutrino fluxes.
We showed that the prompt atmospheric neutrino flux for $E_\nu\sim 10^5-10^7$~GeV comes mostly from the decay of charm hadrons in the CM rapidity range of $4.5 < y < 7.2$. 
Although there are not any planned collider experiments capable of direct measurements of distributions of charm hadrons in this rapidity interval, the range  $\eta_\nu>7.2$ has just started to be probed by the forward neutrino experiments at the LHC, SND@LHC \cite{SNDLHC:2023pun} and FASER$\nu$ \cite{FASER:2023zcr}, which are operating during Run 3 and are sensitive to neutrinos from charm (and other) decays.
This range of $\eta_\nu$ can also be probed during the HL-LHC by the successors of the previous experiment and further new-generation forward experiments in the recently proposed FPF. 
The key observation here is that, 
even if a fraction of
prompt neutrinos in the range of $\eta_\nu>7.2$ at the LHC come from charm hadrons with $y>7.2$, the prompt neutrino fluxes at the FPF experiments are dominated by the decays of charm hadrons with $4.5 < y < 7.2$ as shown in the left panel of figure \ref{fig:dsdeta2}, for the neutrino energies shown in the left panel of figure \ref{fig:dsdE2}. The kinematic overlap with the region presently most interesting for prompt neutrino fluxes emphasizes the deep connection between atmospheric prompt neutrino fluxes and neutrino production from heavy-flavour at the FPF. 

At present, there are no direct LHC measurements of charm production in $pp$ collisions beyond the LHCb coverage that extends up to a CM rapidity $y \le 4.5$. 
LHCb has also presented some analyses in fixed-target modality, using one of the LHC beams  on a gaseous nuclear target, such as He, Ne and Ar.  ALICE has also made very preliminary studies concerning an upgrade for $pA$ measurements in fixed-target modality, using nuclei different from those considered by LHCb. 
In fixed-target modality the LHCb-SMOG2 apparatus, active during Run 3, has a collider-frame rapidity coverage $-2.8 \lesssim  y \lesssim 0.2$, whereas the ALICE fixed-target extension at HL-LHC will cover rapidities $-3.6 \lesssim y \lesssim -2.6$~\cite{Hadjidakis:2019vpg}. 
These experiments have a nucleon-nucleon $\sqrt{s}$ reach $\mathcal{O} (100)$ GeV, which is of limited interest as for directly probing prompt atmospheric neutrino production, considering the $\sqrt{s}$ relevant for the latter. They are however useful to probe the large-$x$ PDF behaviour, with impact even on the forward high-energy prompt neutrino component at the LHC.

In synergy with existing and forthcoming LHCb measurements of charm hadrons with $2.0<y<4.5$, FPF measurements of high-rapidity neutrinos from heavy flavour will help pin down the small-$x$ gluon PDF uncertainty \cite{Anchordoqui:2021ghd,Feng:2022inv} 
and possibly reveal signs of saturation effects, at a scale that has been difficult to establish. 
The gluon PDF uncertainties are currently a subject of intense discussion, as demonstrated by the fact that the range of predictions of different central PDFs and the PROSA PDF uncertainty band do not fully overlap, as shown in figure~\ref{fig:e3phi-x}.  We emphasize that pinning down PDF uncertainties is not only relevant for reducing the present uncertainties on prompt neutrino fluxes at high energy (see section 4), but also to allow an ambitious precision physics program at the HL-LHC and at future higher-energy hadron colliders. We note, however, that our conclusions about the charm hadron rapidity range relevant to prompt atmospheric neutrinos with $E_\nu=10^5-10^7$~GeV are consistent with all of the PDF sets considered here.

While not emphasized here, renormalization and factorization scale uncertainties are large for predictions of both the prompt neutrinos at the neutrino detectors of the FPF~\cite{Bai:2020ukz,Bai:2021ira}  and of the prompt atmospheric neutrino fluxes~\cite{Enberg:2008te,Garzelli:2016xmx,Benzke:2017yjn, Zenaiev:2019ktw, 
Ostapchenko:2022thy,Gauld:2015kvh,Bhattacharya:2015jpa,Bhattacharya:2016jce}. The inclusion of higher order QCD corrections will narrow down the scale uncertainties.
FPF neutrino measurements at high energies will help testing the robustness of the perturbative QCD description of charm production and the importance of non-perturbative effects.   

For prompt atmospheric neutrinos, the focus is on the $\nu_\mu + \bar{\nu}_\mu$ flux (and the essentially equal flux of $\nu_e + \bar{\nu}_e$). The $\nu_\tau + \bar{\nu}_\tau$ atmospheric flux is about an order of magnitude smaller than the $\nu_\mu + \bar{\nu}_\mu$ atmospheric flux \cite{Bhattacharya:2016jce}.   
On the other hand,
at the FPF, the $\nu_\mu + \bar{\nu}_\mu$ and $\nu_e + \bar{\nu}_e$ fluxes have large contributions from pions and kaons, respectively \cite{Feng:2022inv,Kling:2021gos}. However, neutrinos of the highest energies for these two flavours at the FPF have substantial contributions from charm. 
Additionally, almost all the $\nu_\tau + \bar{\nu}_\tau$  come from prompt charm decays. Tau neutrino and antineutrino measurements at the FPF, in turn, would help untangle charm contributions from pion and kaon contributions to other neutrino flavours.  A better understanding of high-energy, forward production of pions and kaons will inform modeling 
of conventional atmospheric neutrino fluxes 
and of extensive air showers produced by cosmic rays and may help solving outstanding issues associated with the current discrepancy between the number of muons detected and predicted at Earth \cite{Albrecht:2021cxw}. At high energies, since 
the prompt diffuse flux of atmospheric muons
is nearly equal to the prompt flux of muon neutrinos \cite{Gondolo:1995fq}, even measurements of the atmospheric muon flux \cite{IceCube:2015wro,Fuchs:2017nuo,Soldin:2018vak} may be of some utility for constraining theoretical predictions of neutrino fluxes from charm.

FCC running in hadron-hadron collision mode with $\sqrt{s}=100$ TeV would directly connect to prompt atmospheric neutrinos with energies $E_\nu = 10^7~\gev$ and beyond.
If the FCC employs experiments that probe the relevant charmed hadron rapidity region for prompt atmospheric neutrino searches (the higher $E_\nu$, the larger are the relevant rapidities, see figure~9), it can provide 
unprecedented information for the prompt flux of atmospheric neutrinos at the highest energies. 
Measurements at the FCC might be relevant for future generation neutrino telescopes, expected to extend the statistics of high-energy neutrino events for neutrino energies beyond the PeV. 
On the other hand, importantly, we emphasize that LHC energies are high enough to probe prompt neutrino production in the current domain of present very large volume neutrino telescopes and in the region where the prompt atmospheric neutrino fluxes are expected to overcome the conventional ones, considering that prompt neutrino production up to $E_\nu \sim \mathcal{O}$(PeV) is dominated by nucleon-nucleon collisions within LHC maximum $\sqrt{s}$ value. 
Measurements of prompt neutrinos at forward neutrino experiments at the LHC through Run 3 and HL-LHC stages will shed light on the nucleon structure 
and will definitely impact the predictions of prompt atmospheric neutrino fluxes.

%\newpage

\acknowledgments
This work is supported in part by U.S. Department of Energy Grant DE-SC-0010113 and DE-SC-0012704 and the National Research Foundation of Korea (NRF) grant funded by the Korea government Ministry of Science and ICT (MSIT) %\footnote{MSIT : Ministry of Science and ICT}) 
No. 2021R1A2C1009296. The work of M.V.G. is supported in part by the Bundesministerium f\"ur Bildung und Forschung under contract 05H21GUCCA.

\bibliography{AtmNu}

\end{document}